\documentclass{elsarticle}
\usepackage{graphicx,float,rotating,multirow,url,natbib}
\usepackage{amsfonts,amssymb,amsmath,latexsym,color,pifont,epsfig,mathdots,mathtools}
\usepackage{setspace,subfigure,multicol,color,xspace,pdfwidgets}

\newcommand{\p}{\partial}
\newcommand{\tht}{\theta}
\newcommand{\om}{\omega}

\begin{document}
\title{Wave-induced dynamics of flexible blades}
\author{M.~Luhar\corref{cor1}\fnref{fn1}}
\ead{luhar@usc.edu}
\author{H.M.~Nepf\corref{cor2}}
\ead{hmnepf@mit.edu}

\address{Department of Civil and Environmental Engineering, Massachusetts Institute of Technology, Cambridge, MA}
\cortext[cor1]{Corresponding author}
\fntext[fn1]{Current address: Department of Aerospace and Mechanical Engineering, University of Southern California, Los Angeles, California, CA}

\begin{abstract}
In this paper, we present an experimental and numerical study that describes the motion of flexible blades, scaled to be dynamically similar to natural aquatic vegetation, forced by wave-induced oscillatory flows.  For the conditions tested, blade motion is governed primarily by two dimensionless variables: (i) the Cauchy number, $Ca$, which represents the ratio of the hydrodynamic forcing to the restoring force due to blade stiffness, and (ii) the ratio of the blade length to the wave orbital excursion, $L$.  For flexible blades with $Ca \gg 1$, the relationship between drag and velocity can be described by two different scaling laws at the large- and small-excursion limits.  For large excursions ($L \ll 1$), the flow resembles a unidirectional current and the scaling laws developed for steady-flow reconfiguration studies hold.  For small excursions ($L \gg 1$), the beam equations may be linearized and a different scaling law for drag applies.  The experimental force measurements suggest that the small-excursion scaling applies even for intermediate cases with $L \sim O(1)$.  The numerical model employs the well-known Morison force formulation, and adequately reproduces the observed blade dynamics and measured hydrodynamic forces without the use of any fitted parameters.  For $Ca \gg 1$, the movement of the flexible blades reduces the measured and modeled hydrodynamic drag relative to a rigid blade of the same morphology.  However, in some cases with $Ca \sim O(1)$, the measured hydrodynamic forces generated by the flexible blades exceed those generated by rigid blades, but this is not reproduced in the model.  Observations of blade motion suggest that this unusual behavior is related to an unsteady vortex shedding event, which the simple numerical model cannot reproduce.  Finally, we also discuss implications for the modeling of wave energy dissipation over canopies of natural aquatic vegetation.
\end{abstract}


\maketitle

\section{Introduction}\label{sec:introduction}
From salt-marshes to seagrass beds and kelp forests, flexible vegetation is ubiquitous in wave-dominated coastal zones.  In all of these systems, the physical interaction between the wave-induced flow and the plants plays an important role in mediating geomorphological, biological and chemical processes.  For instance, the drag generated by the plants leads to a dissipation of wave energy and a damping of the near-bed flow \citep{Kobayashi1993,Lowe2005,Luhar2010}, which inhibits sediment suspension and transport.  The resulting low-flow environment serves as habitat for many species of fish, shellfish and other aquatic organisms.  The physical fluid-structure interaction also determines the posture and motion of the plants.  In addition to influencing light availability \citep{Zimmerman2003}, plant posture and motion mediate nutrient uptake \citep{Hurd2000,Huang2011} and oxygen efflux \citep{Mass2010}, which are some of the most important ecological services provided by aquatic vegetation \citep{Costanza1997}.

Because of its importance to coastal protection, wave energy dissipation over aquatic vegetation has received significant attention in the literature.  It has been studied in the laboratory \citep{Fonseca1992, Kobayashi1993,Manca2012}, in the field~\cite{Bradley2009,Infantes2012}, and using analytical methods or numerical models \citep{Kobayashi1993, Mendez1999, Mendez2004}.  Unfortunately, an accurate prediction of wave attenuation is complicated by the fact that it requires knowledge of the dynamics of the individual elements as well as the interactions between neighboring plants.  It is widely recognized that the rate of energy dissipation at the scale of individual plants depends on the relative motion between the fluid and the vegetation.  Yet, there is no universally-accepted methodology to predict or account for vegetation motion.  As a result, most studies thus far have been restricted to employing bulk drag coefficients that are calibrated to the observations \citep[see e.g.][]{Mendez2004,Bradley2009,Manca2012}.

Several papers have proposed simple models for vegetation motion under wave-forcing to predict the hydrodynamic forces experienced by the plants and quantify the rate of wave energy dissipation.  For example, \citet{Mendez1999} assumed that the flexible vegetation can be modeled as flat stems hinged at the base (i.e. linearly varying deflection with height) whose motion depends on the hydrodynamic forces.  \citet{Mullarney2010} developed an analytical model to predict the motion of single-stem vegetation, showing that vegetation motion depends on a dimensionless parameter representing the ratio of the hydrodynamic drag and vegetation stiffness.  This analytical model accounts for plant motion and morphological variations along the entire length of the stem.  However, it is also restricted to small stem deflections (i.e. linearized Euler-Bernoulli beam theory), and does not include the effects of vegetation buoyancy or inertial forces such as added mass.  In a recent numerical and experimental study, \citet{Zeller2014} developed a more complete model capable of simulating finite-amplitude deflections while accounting for drag as well as added mass.  This effort indicated that the drag generated by the vegetation depends strongly on the ratio of the blade tip excursion to the wave orbital excursion.  Recognizing that this ratio is not a practical predictive tool since it requires knowledge of the blade motion, \citet{Zeller2014} also developed a simple algebraic model that was fitted to numerical results in order to predict wave attenuation.

The purpose of the present study is to build on these previous efforts and improve our understanding of the wave-induced dynamic of flexible blades, with the ultimate goal being a simple, predictive framework to account for blade motion in wave energy dissipation models.  Like \citet{Zeller2014}, we pursue a combination of numerical modeling and laboratory experiments.  For simplicity, both the modeling and experimental efforts focus on flexible blades with uniform rectangular cross-sections, characteristic of seagrasses.  However, the model can easily be extended to account for more complex geometries and spatial variations in material properties.  The numerical model is based on the well known Morison force formulation \citep[see e.g.][]{Denny1998} and allows for large blade deflections (\S\ref{sec:model}).  The experiments simultaneously measured the total hydrodynamic force exerted on the blade, imaged blade posture, and measured the local velocity field using particle image velocimetry.  Two different blade materials, four different blade lengths, and eight different wave conditions were tested to yield a total of 64 experimental cases.  These cases were chosen to correspond to environmentally-relevant dimensionless ranges.  Despite the obvious simplification in modeling the hydrodynamic forces acting on the blade, the numerical model reproduces experimental force measurements and blade posture observations with reasonable fidelity, without the use of any fitting parameters.

Importantly, the numerical model also guides scaling analyses (\S\ref{sec:scaling}) that generalize recent advances in our understanding of the reconfiguration of flexible vegetation in steady, unidirectional flows \citep[e.g.][]{Alben2002,deLangre2008,Gosselin2010,Luhar2011} and the wave-induced motion of flexible vegetation at the small-deflection limit \citep{Mullarney2010}.  For unidirectional flows, \citet{Luhar2011} showed that the reconfiguration of aquatic vegetation depends on two dimensionless parameters: the Cauchy number $Ca$, which represents the relative magnitude of the hydrodynamic forcing and the restoring effect of vegetation stiffness, and the buoyancy parameter $B$, which is the ratio of the restoring forces due to buoyancy and stiffness.  For wave-induced oscillatory flows, a few additional parameters also play a role.  These include the Keulegan-Carpenter number $KC$, which represents the ratio of the inertial forces to drag, as well as the ratio of the blade length to the wave orbital excursion, $L$.  As will be shown later, the latter parameter sets the transition between the small deflection limit described by \citet{Mullarney2010} and a quasi-steady situation resembling unidirectional flows, thereby playing a key role in dictating blade behavior.  To characterize the reduction in hydrodynamic forces (and therefore, wave energy dissipation) due to blade flexibility and motion, we propose the use of an effective blade length, defined as the length of a rigid upright blade that generates the same forces as the flexible blade.

\section{Theory}\label{sec:theory}
\subsection{Dynamic blade model}\label{sec:model}
\begin{figure}
	\centering{\includegraphics[scale=0.09]{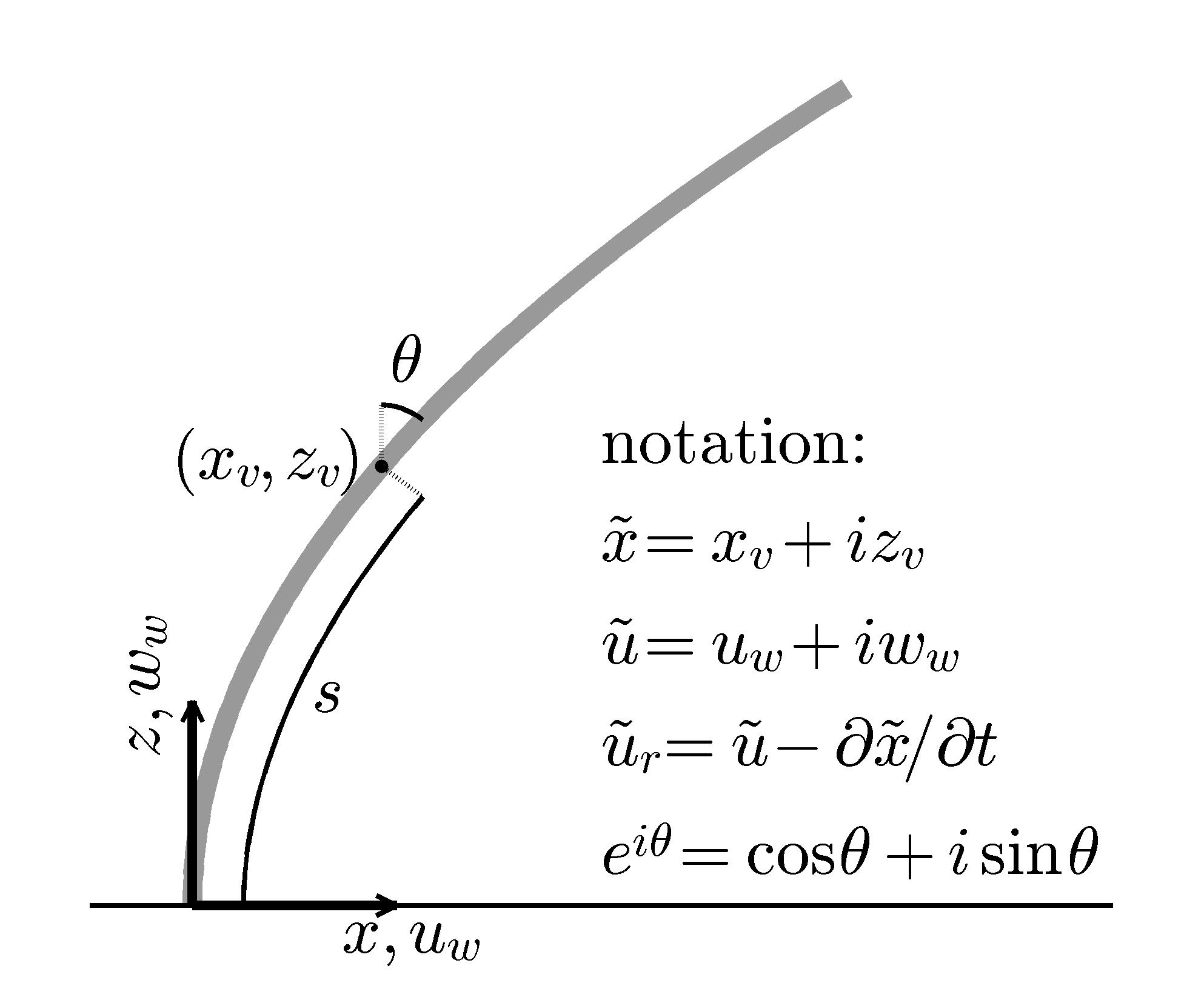}}
	\caption{Schematic showing the coordinate system and notation used for the dynamic blade model.}
	\label{fig:coord}
\end{figure}

The model considers inextensible blades of width $b$, thickness $d$, length $l$, elastic modulus $E$, and density $\rho_v$ moving in a two-dimensional plane.  The blade is assumed to move without twisting, such that the frontal area exposed to the flow is always $b$ per unit blade length.  The coordinate system used is shown in Fig.~\ref{fig:coord}, where $s$ is the distance along the blade from the base and $\tht$ is the local bending angle of the blade relative to the vertical.  The flow field is described by horizontal and vertical velocities that vary in time ($t$) as well as the vertical ($z$) direction: $u_w(z,t)$ and $w_w(z,t)$, respectively.  In general, the velocity fields considered in this paper are oscillatory with period $T_w$.  The exact forms of $u_w$ and $w_w$ used to force the model are discussed in \S\ref{sec:methods}.  Complex notation is used to describe the force balance for the blade in both the horizontal and vertical directions, such that the velocity is $\tilde{u}=u_w+i w_w$, and $\tilde{x} = x_v(s) + i z_v(s)$ describes the position of a point along the blade in the $x-z$ (horizontal-vertical) plane.  Using the standard definition $e^{i\tht}=\cos\,\tht+i\sin\,\tht$, the inextensibility condition can be expressed as:

\begin{equation}\label{eq:xztheta}
\tilde{x} =\int\limits_{0}^{s}i\exp(-i\tht)\,ds^\prime =\int\limits_{0}^{s}\sin\tht \,ds^\prime + i\int\limits_{0}^{s}\cos\tht \,ds^\prime
\end{equation}

{\noindent}where $ s^\prime$ is a dummy variable.  The relative velocity between the blade and the fluid is defined as $\tilde{u}_r = \tilde{u}-\p\tilde{x}/\p t$ (Fig.~\ref{fig:coord}).

The dynamics of the blade are controlled by a number of internal and external forces.  The internal forces include the tension $T$, which acts in the blade-parallel direction, and the shear $V=-EI(\p^2\tht/\p s^2)$, which acts in the blade-normal direction.  Here, $I$ is the second moment of area for the blade cross-section ($I=bd^3/12$ for rectangular cross-sections).  The external forces acting on the blade are assumed to be described by the well-known Morison formulation \citep{Denny1998}.  The external forces per unit blade length include (i) the net buoyancy force

\begin{equation}\label{eq:fb}
f_B = (\rho-\rho_v)gbd,
\end{equation}

{\noindent}which acts in the vertical direction, (ii) the Froude-Krylov (or virtual buoyancy) force arising from the unsteady pressure gradient in the fluid \citep{Batchelor2000}

\begin{equation}\label{eq:fvb}
f_{VB} = \rho bd \frac{\p\tilde{u}}{\p t},
\end{equation}

{\noindent}which acts in the direction of flow acceleration, (iii) the drag force

\begin{equation}\label{eq:fd}
f_D = \frac{1}{2}\rho C_D b \left|\Re\left(\tilde{u}_r e^{i\tht}\right)\right|\Re\left(\tilde{u}_r e^{i\tht}\right),
\end{equation}

{\noindent}which depends on the relative velocity normal to the blade $\Re(\tilde{u}_r e^{i\tht})$, and acts in the blade-normal direction ($\Re()$ denotes the real component of a quantity and $\Im()$ denotes the imaginary component), (iv) skin friction $f_F$, 

\begin{equation}\label{eq:ff}
f_F = \frac{1}{2}\rho C_F b \left|\Im\left(\tilde{u}_r e^{i\tht}\right)\right|\Im\left(\tilde{u}_r e^{i\tht}\right),
\end{equation}

{\noindent}which acts in the blade-parallel direction and depends on the relative velocity along the blade $\Im(\tilde{u}_r e^{i\tht})$, and finally (v) the added mass force

\begin{equation}\label{eq:fam}
f_{AM} = \frac{\pi}{4}\rho C_M b^2 \Re\left(\frac{\p\tilde{u}_r}{\p t} e^{i\tht}\right),
\end{equation}

{\noindent}which depends on the relative acceleration between the flow and the blade, and also acts in the blade-normal direction.  Here, $\rho$ is the density of water, $C_D$ is the drag coefficient, $C_F$ is the skin friction coefficient, and $C_M$ is the added mass coefficient.  Following \citet{Keulegan1958}, the cylinder-equivalent blade cross-sectional area $\pi b^2/4$ is used to represent the added mass force.

A balance of the internal and external forces described above yields the following physically-intuitive equation governing blade motion:

\begin{equation}\label{eq:governing}
\frac{\p}{\p s}\left((V+iT)e^{-i\tht}\right) + i f_B + (f_D + if_F + f_{AM})e^{-i\tht} + f_{VB} 
= \rho_v bd \frac{\p^2\tilde{x}}{\p t^2}.
\end{equation} 

{\noindent}The term on the right-hand side of Eq.~\ref{eq:governing} represents blade inertia.  The real part of Eq.~\ref{eq:governing} represents the horizontal force balance, and the imaginary part represents the vertical force balance.  Note that the blade-normal ($V$, $f_D$, $f_{AM}$) and blade-parallel ($T$, $f_F$) forces have been multiplied by the factor $e^{-i\tht}$ to rotate them into the $x-z$ directions.  Expanding the first term on the left-hand side of Eq.~\ref{eq:governing}, introducing the expression for $f_{VB}$ shown in Eq.~\ref{eq:fvb}, and multiplying by $e^{i\tht}$ yields:

\begin{eqnarray}\label{eq:governing2}
&& \frac{\p}{\p s}\left(V+iT\right) -i\frac{\p\tht}{\p s}\left(V+iT\right)\nonumber \\
&& \nonumber \\
&& +if_Be^{i\tht} + f_D + i f_F + f_{AM}
+\rho bd\left(\frac{\p \tilde{u}}{\p t} -\frac{\rho_v}{\rho}\frac{\p^2\tilde{x}}{\p t^2}\right)e^{i\tht} = 0
\end{eqnarray}

{\noindent}Multiplication by $e^{i\tht}$ rotates the force balance in Eq.~\ref{eq:governing} such that the real part of Eq.~\ref{eq:governing2} represents the blade-normal force balance, and the imaginary part represents the blade-parallel force balance.

To make the governing Eq.~\ref{eq:governing2} dimensionless, the following normalized variables are used (denoted by over-hats):

\begin{equation}
s=l\hat{s}\;\; ; \;\;t=\hat{t}/\om\;\; ; \;\;\tilde{u}=U_w\hat{u}\;\; ; \;\;T =(EI/l^2)\hat{T}\;\; ; \;\;\tilde{x} = l\hat{x}
\end{equation}

{\noindent}The blade coordinates $ s$ and $\tilde{x}$ are normalized by the blade length, $l$.  Time is normalized by the wave radian frequency, $\om = 2\pi/T_w$.  Velocity is normalized using the horizontal oscillatory velocity scale, $U_w = A_w \om$, where $A_w$ is the horizontal wave excursion.  Tension is normalized with the assumed scaling for the internal shear force, $EI/l^2$.  Substituting these normalized variables, along with the expressions for $V$ and the external forces (Eqs.~\ref{eq:fb}-\ref{eq:fam}) in Eq.~\ref{eq:governing2}, and dividing through by the factor $EI/l^3$, yields the following dimensionless equation describing blade dynamics:

\begin{eqnarray}\label{eq:governdim}
&&\frac{\p}{\p\hat{s}}\left(-\frac{\p^2\tht}{\p\hat{s}^2}+i\hat{T}\right)
-i\frac{\p\tht}{\p \hat{s}}\left(-\frac{\p^2\tht}{\p\hat{s}^2}+i\hat{T}\right) 
\nonumber\\
&&\nonumber\\
&&+iBe^{i\tht}
+\frac{1}{2}C_D Ca |\Re(\hat{u}_r e^{i\tht})|\Re(\hat{u}_r e^{i\tht}) 
+i \frac{1}{2}C_F Ca |\Im(\hat{u}_r e^{i\tht})|\Im(\hat{u}_r e^{i\tht}) 
\nonumber\\
&&\nonumber\\
&&+\frac{2\pi^2}{4}C_M\frac{Ca}{KC}\Re\left(\frac{\p\hat{u}_r}{\p\hat{t}}e^{i\tht}\right) 
+2\pi \frac{Ca S}{KC}\left(\frac{\p\hat{u}}{\p\hat{t}}-\rho^\prime L\frac{\p^2\hat{x}}{\p\hat{t}^2}\right)e^{i\tht} = 0,
\end{eqnarray}

{\noindent}in which

\begin{equation}\label{eq:ur}
\hat{u}_r = \hat{u}-L\left({\p\hat{x}}/{\p\hat{t}}\right)
\end{equation} 

{\noindent}is the dimensionless relative velocity between the blade and the water, and

\begin{equation}\label{eq:L}
L=\frac{l\om}{U_w}=\frac{l}{A_w}
\end{equation}

{\noindent}is the ratio of the blade length to the wave excursion.  The dimensionless parameters governing blade motion include the Cauchy number ($Ca$) and the buoyancy parameter ($B$):

\begin{equation}\label{eq:Ca}
Ca = \frac{\rho bU_w^2l^3}{EI}
\end{equation}

\begin{equation}\label{eq:B}
B=\frac{(\rho-\rho_v) gbdl^3}{EI}
\end{equation}

{\noindent}These parameters are similar to those governing the reconfiguration of flexible blades in unidirectional flows \citep{Luhar2011}, but with $U_w$ replacing the uniform current speed.  A number of additional parameters also become important in unsteady oscillatory flow, including the Keulegan-Carpenter number $KC = U_w T_w/b$, which represents the ratio of the drag force to the inertial forces.  Note that, unlike \citet{Luhar2011}, we do not include $C_D$ in the definition of $Ca$ in this paper.  This is because the drag coefficient is known to vary with $KC = U_wT_w/b$ for wave-induced oscillatory flows (see \S \ref{sec:CDCM}).  Finally, Eq.~\ref{eq:governdim} also includes the ratio of densities, $\rho^\prime=\rho_v/\rho$, and the blade slenderness, $S=d/b$.

The boundary conditions for this model are: clamped at the base of the blade $\tht = 0$ at $\hat{s}=0$, and free at the tip $(\p\tht/\p\hat{s})=(\p^2\tht/\p\hat{s}^2)=\hat{T}=0$ at $\hat{s}=1$.  The inextensibility condition (Eq.~\ref{eq:xztheta}) remains the same under the normalization, except that $\tilde{x}$ and $ s$ are replaced by their dimensionless counterparts $\hat{x}$ and $\hat{s}$, respectively.

\subsection{Model coefficients and numerical implementation}\label{sec:CDCM}
The model described in the previous section requires an accurate description of $C_D$ and $C_M$.  \citet{Luhar2011} showed that the flat plate drag coefficient for steady flows, $C_D = 1.95$, can be used to accurately capture the drag generated by flexible blades in unidirectional flow, as long as the blade-normal velocity is used in the quadratic law.  Similarly, we hypothesize that the flat plate $C_D$ and $C_M$ may also be used for flexible blades in oscillatory flows, as long as the blade-normal \textit{relative} velocity and acceleration are used to characterize the drag and added mass forces.

\begin{figure}
	\centering{\includegraphics[scale=0.09]{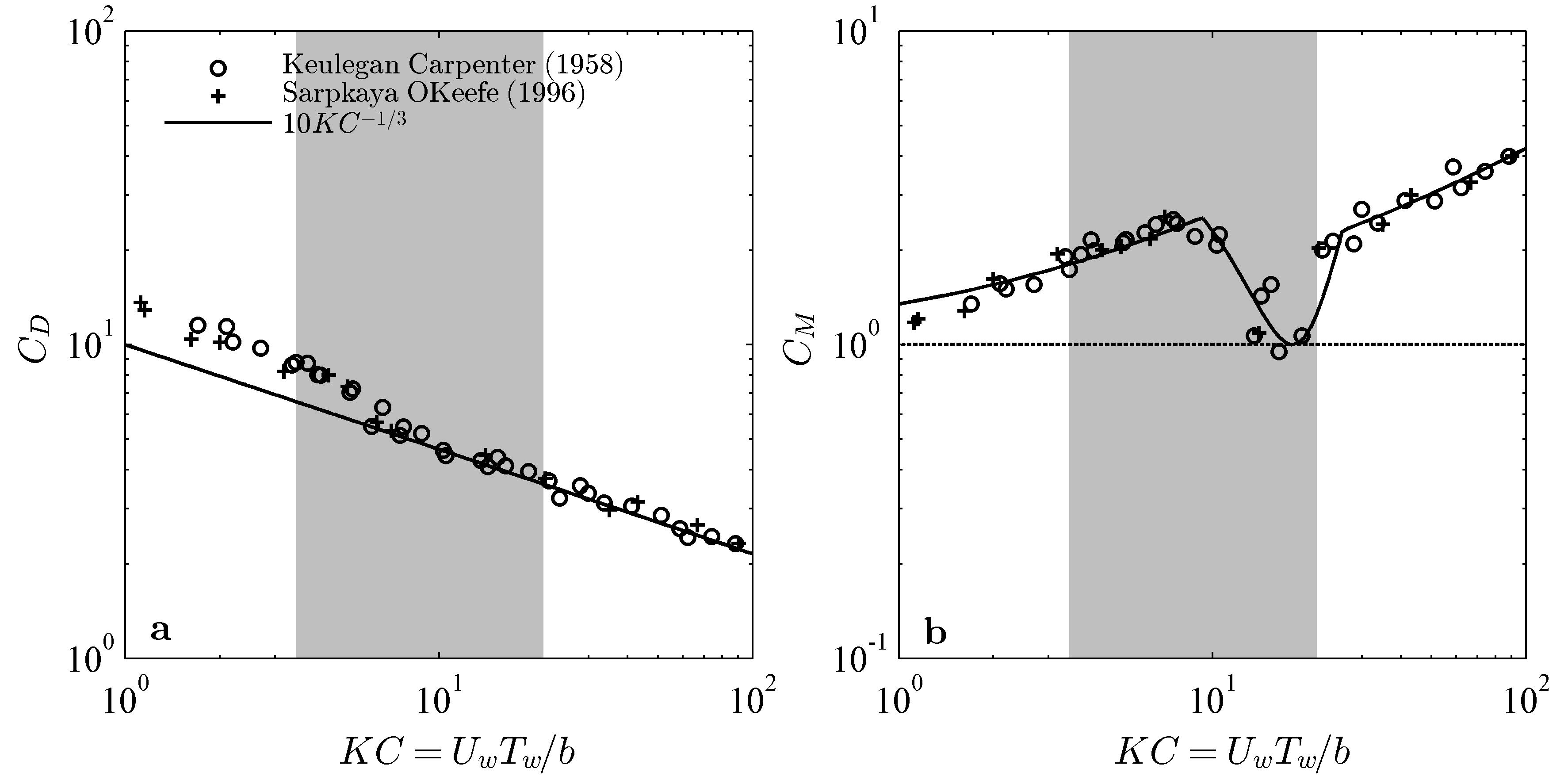}}
	\caption{Drag and added mass coefficients, $C_D$ (a) and $C_M$ (b), for rigid flat plates in oscillatory flows plotted against the Keulegan-Carpenter number, $KC$.  The data shown are from~\cite{Keulegan1958,Sarpkaya1996}.  The shaded regions represent the range of $KC$ for the laboratory experiments described in $\S$\ref{sec:experiments}.}
	\label{fig:CDCM}
\end{figure}

Flat-plate $C_D$ and $C_M$ from previous experiments \citep{Keulegan1958,Sarpkaya1996} are plotted in Fig.~\ref{fig:CDCM}.  Both data sets show that $C_D$ and $C_M$ depend on the Keulegan-Carpenter number, $KC$.  Following Graham \citep{Graham1980}, we model the relationship between the drag coefficient and Keulegan-Carpenter number as $C_D = 10 KC^{-1/3}$ (solid line in Fig.~\ref{fig:CDCM}a).  However, as $KC\to\infty$ (i.e., $T_w\to\infty$), the drag coefficient must approach the steady flow value, $C_D = 1.95$.  Therefore, a more complete definition is $C_D = \max(10 KC^{-1/3}, 1.95)$.  Strictly, the drag coefficient also depends on the Reynolds number $Re = U_w b /\nu$, where $\nu$ is the kinematic viscosity of the fluid.  For flat plates in steady flow, this dependence can be approximated as $C_D = 1.95 + 50/Re$ \citep{Ellington1991}.  $Re \ge 1000$ for the cases considered in this study (Table~\ref{tab:lab}), and so the additional term dependent on Reynolds number is predicted to increase $C_D$ by $<3\%$.  Since the $KC$-dependence discussed above yields a near-$50\%$ decrease in $C_D$ over the conditions tested (shaded region in Fig.~\ref{fig:CDCM}a), the model does not account for the comparatively minor influence of $Re$.

Unlike the monotonically decreasing relationship between $C_D$ and $KC$, the variation of $C_M$ with $KC$ is more complex.  In general, $C_M$ increases gradually with $KC$, but there is a pronounced dip in $C_M$ at $KC\approx 18$.  This dip corresponds to the conditions in which a single eddy is shed from the plate during each wave half-cycle \citep{Keulegan1958}.  We use the spline shown as a solid-line in Fig.~\ref{fig:CDCM}b to describe the variation in $C_M$ with $K_C$.  However, a constant value for the added mass coefficient, $C_M =1$, does not significantly alter the results obtained from the numerical model for all the cases considered here.  The mean ($\pm$ standard deviation) ratio of the maximum forces predicted by the numerical model with $C_M=1$ and the spline shown in Fig.~\ref{fig:CDCM} is $0.93 (\pm 0.08)$.

To predict blade motion and drag, we solve the governing Eq.~\ref{eq:governdim} numerically using a finite difference scheme that is second-order accurate in space.  The blade-normal force balance (real component of Eq.~\ref{eq:governdim}) is solved explicitly to yield $\tht(\hat{s})$.  Although, the third-order spatial derivative is treated implicitly for stability.  Further, the quadratic drag term is linearized by using the magnitude of the blade-normal velocity ($|\Re(\hat{u}_r e^{i\tht})|$) from the previous time step.  The blade-parallel force balance (imaginary component of Eq.~\ref{eq:governdim}) is solved using backward differences to yield the tension $\hat{T}(\hat{s})$ at every time step.  A constant skin friction coefficient $C_F = 0.1$ is employed, which is at the upper end of the range for horizontal flat plates in steady flow \citep{Kundu2004}.  The exact value of the skin friction coefficient did not have an appreciable effect on the results.  Specifically, the ratio of the predicted root-mean-square (RMS) forces for $C_F = 0.1$ and $C_F = 0.01$ was distributed with mean and standard deviation $1.00 \pm 0.01$.  However, some of the simulations at high Cauchy numbers proved to be unstable for the lower value of $C_F = 0.01$.

We carried out numerical simulations corresponding to each of the sixty four cases tested in the laboratory (see Table~\ref{tab:lab} for greater details).  The simulations were performed with a spatial resolution of 512 equally spaced grid points from $\hat{s}=0$ to $\hat{s}=1$ and a temporal resolution of 120 intervals for each wave period.  These resolutions were chosen to minimize computational expense while maintaining stability and fidelity.  As a rough estimate of convergence, the RMS forces predicted using lower resolutions (256 grid points, 60 intervals per period) differed by less than $1\%$; the ratio of the predicted RMS forces at each resolution was distributed with mean $\pm$ standard deviation: $0.99\pm 0.02$.  Note that the spatial and temporal resolution had to be refined together for stability purposes.  The numerical simulations were run until a quasi-steady state was achieved i.e., once blade motion did not vary from one wave cycle to the next.  Typically, this quasi-steady state was achieved within 7 or 8 wave cycles.  Further details on the discretization and a complete code listing can be found in \citep{LuharPhD}.

\subsection{Scaling considerations}\label{sec:scaling}
\subsubsection{Steady flow}
As noted earlier, \citet{Luhar2011} showed that the reconfiguration of flexible blades in steady flows is determined by the Cauchy number $Ca$ (Eq.~\ref{eq:Ca}) and the buoyancy parameter $B$ (Eq.~\ref{eq:B}).  When the hydrodynamic forcing is much smaller than the restoring force due to stiffness $Ca \ll 1$ or buoyancy $Ca \ll B$, the blade remains upright in the flow.  At this effectively-rigid limit, the hydrodynamic drag generated by the blade is predicted well by assuming a flat plate drag coefficient.  However, as the velocity increases such that the hydrodynamic forcing becomes larger than the restoring forces due to stiffness and buoyancy, $Ca > O(1)$ and $Ca > O(B)$, the blade starts to reconfigure, or bend, in the flow.  To quantify the resulting drag reduction, \citet{Luhar2011} proposed the use of an effective blade length $l_e$.  This is defined as the length of a rigid, vertical blade that generates the same horizontal drag as the flexible blade of length $l$.  \citet{Luhar2011} showed that when the hydrodynamic forcing becomes much larger than blade buoyancy and the restoring force due to blade stiffness, $Ca \gg B$ and $Ca \gg 1$, the effective length scales as $l_e/l \sim Ca^{-1/3}$.  This scaling, first reported by \citet{Alben2002} for the case of flexible plates without buoyancy, represents the following balance between the restoring force due to stiffness and drag in the reconfigured state: $EI/l_e^2 \sim \rho b l_e U^2$.  In other words, both the blade curvature, $(\partial^2\theta/\partial s^2)$, and the pressure drag force, $F_x$, scale on the effective length, $l_e$, in the reconfigured state.

\subsubsection{Unsteady flow with large excursions ($L \ll 1$)}
\begin{figure}
	\centering{\includegraphics[scale=0.09]{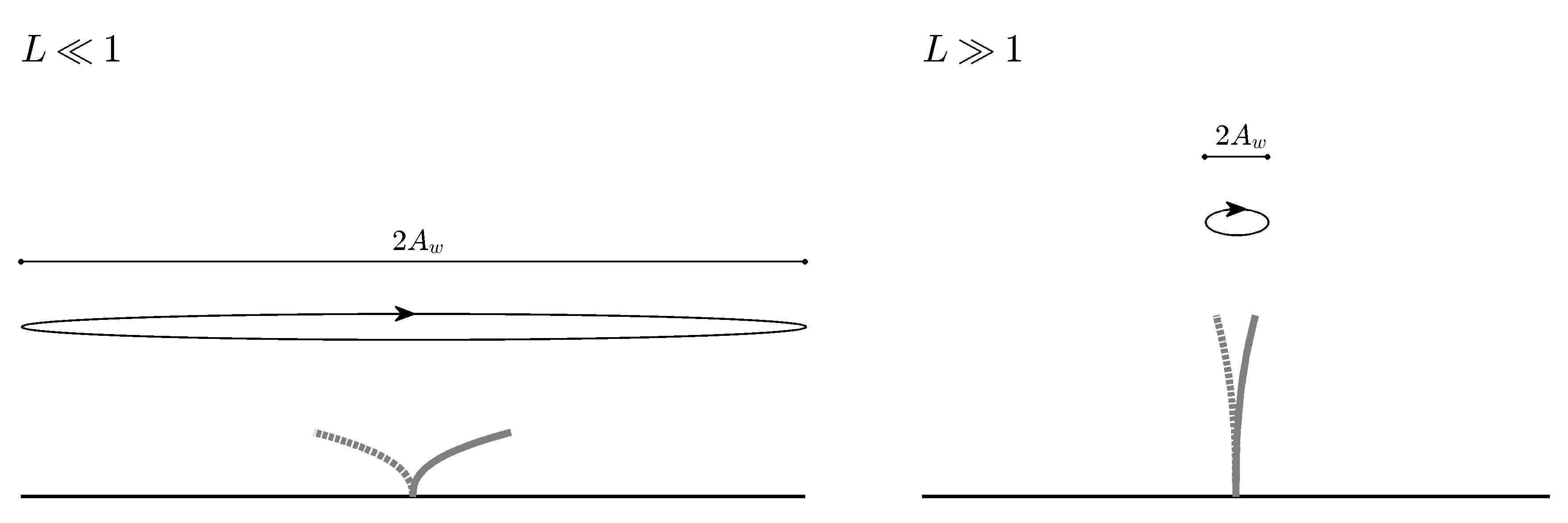}}
	\caption{Schematic illustrating the difference in blade behavior at the limit of large and small wave excursions, $L\ll 1$ and $L \gg 1$, respectively.}
	\label{fig:L}
\end{figure}

For the unsteady case, Eq.~\ref{eq:ur} shows that as the wave excursion becomes much greater than the blade length, $A_w \gg l$ ($L \ll 1$), the relative velocity between the blade and the water is approximately equal to the water velocity, $\hat{u}_r \approx \hat{u}$, over most of the wave cycle.  Since the blade width is typically smaller than the blade length for natural aquatic vegetation such as seagrasses, $b/l < 1$, the Keulegan-Carpenter number is also larger than unity at this large excursion limit, $KC = U_wT_w/b = 2\pi A_w/b \gg 1$.  This suggests that the inertial terms (added mass, virtual buoyancy, blade inertia) in the last row of Eq.~\ref{eq:governdim} can be neglected.  For $\hat{u}_r \approx \hat{u}$ and negligible inertia, Eq.~\ref{eq:governdim} resembles the steady flow reconfiguration model developed by \citet{Luhar2011} and so we expect the results discussed in the previous paragraph to hold.  Specifically, the effective length scales as $l_e/l \sim Ca^{-1/3}$ for $L \ll 1$, $Ca \gg 1$ and $Ca\gg B$, with no dependence on $L$.  Physically, at this large excursion and large-drag limit, we have a quasi-steady situation where a flexible blade can be pushed over by the flow in the early stages of a wave-half cycle (see Fig.~\ref{fig:L}).  The blade remains bent until the oscillatory flow reverses direction at the end of the wave half-cycle, with the bent posture reflecting a balance between hydrodynamic drag and the restoring forces due to buoyancy and stiffness.

\subsubsection{Unsteady flow with small excursions ($L \gg 1$)}
At the other high $Ca$-limit, for which $L \gg 1$, the horizontal wave excursion is much smaller than the blade length, $A_w \ll l$.  In this case, we anticipate that the blade remains nearly vertical as it moves back and forth throughout the wave cycle (see Fig.~\ref{fig:L}), and that the horizontal blade excursion scales with the wave excursion, i.e., $|x_v|\sim O(A_w)$.  More formally, we expect that $\tht\sim O(L^{-1}) \ll 1$ and so the inextensibility condition becomes:

\begin{equation}\label{eq:xzlin}
\hat{x}=\int\limits_{0}^{\hat{s}}i\exp(-i\tht)\,d\hat{s}^\prime \approx \int\limits_{0}^{\hat{s}}i(1-i\tht)\,d\hat{s}^\prime = i\hat{s}+\int\limits_{0}^{\hat{s}}\tht\,d\hat{s}^\prime
\end{equation}

{\noindent}In dimensional terms, this leads to $z_v(s)\approx s$ (i.e., nearly-vertical blade) and $(\p x_v/\p s) \approx \tht$ (i.e., $|x_v|\sim \tht l \sim A_w$).  At this small-deflection limit, the blade curvature term can be linearized such that $EI(\partial^2\theta/\partial s^2) \approx EI(\partial^3 x_v/ \partial z_v^3)$ (see Eq.~\ref{eq:xzlin}).  Since the blade excursion scales on the wave excursion, balancing drag and blade stiffness for this limit of $L \gg 1$ yields:

\begin{equation}\label{eq:fbalancele}
EI\frac{\partial^3 x_v}{\partial z_v^3} \sim F_x \; \to \; EI \frac{A_w}{l_e^3}\sim \rho b l_e U_w^2
\end{equation}

{\noindent}Using the definition of the Cauchy number $Ca$ (Eq.~\ref{eq:Ca}) and the ratio $L$ (Eq.~\ref{eq:L}), the above equation can be rewritten as:

\begin{equation}\label{eq:lescaling}
({l_e}/{l}) \sim (Ca L)^{-1/4}
\end{equation}

{\noindent}Essentially, with this scaling, the effective length represents the blade length over which there is significant relative motion between the blade and the water.  The upper part of the blade, $z_v > l_e$, moves nearly passively with the flow and therefore, forces are generated primarily in the lower part, $z_v < l_e$.  Note that this small-deflection behavior is identical to that described by the analytical model developed in \citet{Mullarney2010}.  Specifically, \citet{Mullarney2010} showed that for very flexible single-stemmed aquatic vegetation, energy dissipation is concentrated in a thin near-bed elastic boundary layer where the motion of the plants reduces smoothly to zero.  The height of this elastic boundary layer scales as $S_{MH}^{1/4}$, where $S_{MH} \propto (Ca L)^{-1}$ is the dimensionless ratio of stiffness to the hydrodynamic forcing for small blade excursions.  The effective length shown in Eq.~\ref{eq:lescaling} is therefore proportional to the height of this elastic boundary layer, where we expect significant relative motion between the water flow and the vegetation.

Importantly, Eq.~\ref{eq:lescaling} assumes that drag is the dominant hydrodynamic forcing.  This is reasonable for the range of conditions tested in the laboratory, where the Keulegan-Carpenter number, $KC \ge 3.7$.  For $KC \ll 1$, Eq.~\ref{eq:governdim} suggests that the added mass force, which scales as $Ca/KC$, would become the dominant hydrodynamic forcing instead of drag.  At this inertia-dominated limit, a force balance similar to the one shown in Eq.~\ref{eq:fbalancele}, but with added mass instead of drag suggests that $l_e/l \sim (Ca L/KC)^{-1/4}$.

\section{Laboratory experiments}\label{sec:experiments}
\subsection{Materials and Methods}\label{sec:methods}
We pursued laboratory experiments that simultaneously (i) measured the hydrodynamic force generated by flexible blades over a wave-cycle, (ii) imaged blade motion, and (iii) measured the local velocity field using particle image velocimetry (PIV).  The experiments were carried out in a 24 m-long, 38 cm-wide, 60 cm-deep wave flume fitted with a paddle wavemaker.  The paddle was actuated using a programmable signal generator using the waveform suggested in \citet{Madsen1971}.

We tested model blades made of two different materials: silicon foam ($E = 500$ kPa; $\rho_v = 670$ kg m$^{-3}$; $d = 1.9$ mm) and high-density polyethylene (HDPE, $E = 0.93$ GPa; $\rho_v = 950$ kg m$^{-3}$; $d = 0.4$ mm).  The blade width was $b = 2.0$ cm in all cases and the blade length was varied from $l = 5$ cm to $l = 20$ cm in 5 cm increments.  The model blades made of HDPE exhibited a small degree of curvature in the cross section.  Because of this curvature, the HDPE blades had a second moment of area, $I\approx bd^3/6$.  So, the HDPE blades were twice as stiff as they would have been if the cross-section had been perfectly flat and rectangular ($I = bd^3/12$).  This increase in stiffness was confirmed with simple cantilever bending tests.  Specifically, HDPE blades of length 7.5 cm to 15 cm were clamped to be horizontal at one end and allowed to bend under self-weight in air.  Images of the blades were compared to predictions based on a static, nonlinear beam bending equation making the same structural assumptions as Eq.~\ref{eq:governing}-\ref{eq:governdim} (i.e. linearly elastic, constant cross-section, inextensible, finite deformations) for $I = bd^3/4$ to $I = bd^3/10$.  The predictions based on $I = bd^3/6$ provided the best fit and the uncertainty associated with this fit was estimated to be $\Delta I \approx bd^3/30$ (i.e. 20\%).  Note that this assumption of an enhanced stiffness due to curvature is valid for small deformations.  However, the cross-section can flatten when the blade is bent significantly, resulting in a local decrease in the second moment of area $I$.  While a constant $I$ yielded accurate blade bending predictions for the cantilever tests, we do not have any measurements that can provide insight into the importance of these nonlinear three-dimensional effects for the dynamic blade experiments.  Such effects cannot be reproduced in the simple numerical model either.

Each model blade was tested in eight different wave conditions: waves of frequency $f=0.5$ Hz ($T_w = 2.0$ s) with nominal amplitudes $a_w\approx 1, 2, 3, 4$ cm; waves of frequency $f = 0.7$ Hz ($T_w = 1.4$ s) with amplitudes $a_w \approx 2, 4$ cm; waves of frequency $f=0.9$ Hz ($T_w = 1.1$ s) with amplitudes $a_w \approx 2, 4$ cm.  A list of all the test cases is shown in Table~\ref{tab:lab}.

\begin{figure}
	\centering{\includegraphics[scale=0.09]{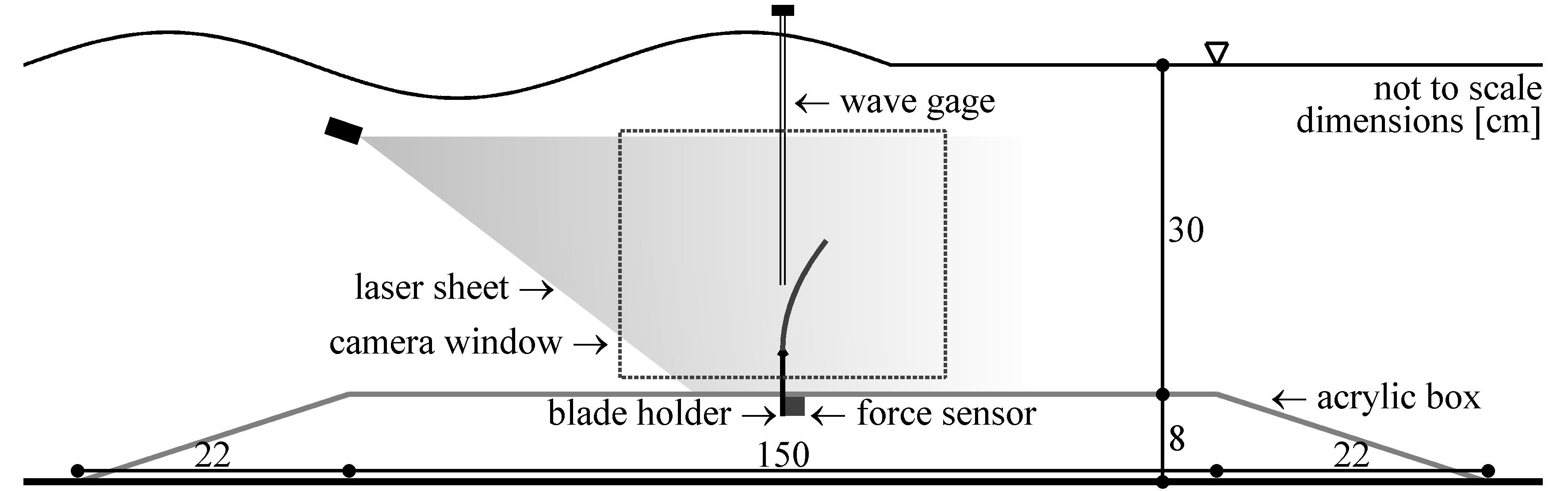}}
	\caption{Schematic showing the experimental setup.  The laser light sheet was placed 0.5 mm behind the model blades.  The wave gage was placed 15 cm behind the blades.  The direction of wave propagation was from left to right.  Not to scale.}
	\label{fig:lab}
\end{figure}

\begin{sidewaystable}[p]
	\centering
	\caption{List of test cases for the dynamic blade experiments.}
	\begin{tabular}{lllllllllll}
	\hline
	& $a_w$ [cm]$^\star$ &               & 1 & 2 & 3 & 4 & 2 & 4 & 2 & 4 \\
	&                    & ($\pm$0.1 cm) & (0.9) & (1.9) & (2.9) & (3.9) & (1.7) & (3.5) & (1.7) & (3.6) \\
	& $T_w$ [s] & ($\pm$0.1 s) & 2 & 2 & 2 & 2 & 1.4 & 1.4 & 1.1 & 1.1 \\
	& $U_w$ [cm s$^{-1}$] & ($\pm$0.9 cm s$^{-1}$) & 5.0 & 10.1 & 15.4 & 20.6 & 8.9 & 16.7 & 6.6 & 12.8 \\
	& $A_w$ [cm] & ($\pm$0.1 cm) & 1.6 & 3.2 & 4.9 & 6.6 & 2.0 & 3.8 & 1.2 & 2.3 \\
	& $KC$ & & 5.0 & 10.1 & 15.4 & 20.6 & 6.4 & 11.9 & 3.7 & 7.1 \\
	\hline
	& $l$ [cm] & $B$ & \multicolumn{8}{c}{$Ca$} \\
	\hline
	HDPE & & & & & & & & & & \\
	$E = 0.93\pm 0.08$ GPa & 5 & 0.002 & 0.02 & 0.12 & 0.28 & 0.50 & 0.09 & 0.36 & 0.06 & 0.20 \\
	$\Delta\rho = 50\pm 10$ kg m$^{-3}$ & 10 & 0.02 & 0.24 & 1.0 & 2.5 & 4.0 & 0.76 & 2.8 & 0.41 & 1.5 \\
	$b = 2.0\pm 0.05$ cm & 15 & 0.06 & 0.81 & 3.2 & 7.2 & 13 & 2.7 & 9.7 & 1.5 & 5.7 \\
	$d = 0.4\pm 0.04$ mm & 20 & 0.15 & 2.2 & 7.5 & 19 & 36 & 6.2 & 21 & 3.5 & 13 \\
	& & & & & & & & & & \\
	Silicon foam & & & & & & & & & & \\
	$E = 500\pm 60$ kPa & 5 & 2.7 & 1.0 & 4.4 & 9.9 & 17 & 3.5 & 12 & 1.7 & 6.8 \\
	$\Delta\rho = 330\pm 50$ & 10 & 22 & 8.9 & 36 & 90 & 160 & 27 & 95 & 15 & 52 \\
	$b = 2.0\pm 0.05$ cm & 15 & 73 & 33 & 120 & 280 & 530 & 100 & 310 & 53 & 210 \\
	$d = 1.9\pm 0.10$ mm & 20 & 170 & 71 & 300 & 680 & 1200 & 240 & 800 & 110 & 470 \\
	\hline
	\multicolumn{11}{l}{$^\star$ In the text, we refer to each wave condition using these values for the amplitude $a_w$.  Measured $a_w$ are shown}\\
	\multicolumn{11}{l}{in parentheses below.}\\
	\end{tabular}%
	\label{tab:lab}%
\end{sidewaystable}%

To measure the total horizontal force generated by the blade, $F_x$, we used a submersible s-beam load sensor (Futek LSB210).  The measurements were logged to a computer using a bridge completion and data acquisition module (National Instruments NI-USB9237).  Based on a calibration with known weights performed prior to the experiments, the resolution of the load cell was 0.001 N and the accuracy was 10$\%$.  To study how $F_x$ varies over a wave cycle, the local wave elevation, $\eta$, was measured synchronously with $F_x$ using a wave gage of 0.2 mm accuracy.  The analog output from the wave gage was amplified and logged to a computer using an analog-digital converter (National Instruments NI-USB6210).  We measured $F_x$ and $\eta$ for a period of 3 min at a sampling rate of 2000 Hz.  Thus, we captured between 90 and 162 waves, depending on wave frequency.  The measurements were then phase-averaged to yield representative descriptions of $F_x$ and $\eta$ over a single wave cycle.  As shown in Fig.~\ref{fig:lab}, the load cell was mounted inside a trapezoidal box of height 8 cm and total length 192 cm.  The model blade was attached to the load cell via a stainless steel blade holder that protruded through a 1.25 cm-diameter hole in the trapezoidal box.  The blade holder placed the base of the blade 4 cm above the box surface.  The total water depth was 38 cm.  Note that the model blade was mounted in the middle of the flume, while the wave gage was mounted approximately 15 cm to the side of the blade at the same $x$-location (i.e., the wave gage was 4 cm from the flume sidewall).

For the blade motion and PIV measurements, illumination was provided by a laser light sheet.  The light sheet was placed in the $x-z$ plane, 0.5 mm behind the model blade (see Fig.~\ref{fig:lab}).  Images were captured at 60 frames per second (fps) using a monochrome CCD camera (Dalsa Falcon 1.4M100HG) of resolution 1400 pixels $\times$ 1024 pixels.  The field of view was approximately 42 cm $\times$ 31 cm, leading to a spatial resolution of 0.03 cm pixel$^{-1}$.  For each case tested, we captured images over 3 wave cycles, e.g., for waves of period $T_w = 2$ s, we captured 6 s worth of images.  For the PIV measurements, the water was seeded with Pliolite particles (density 1020 kg m$^{-3}$).  PIVlab, a MATLAB software package, was used to calculate the horizontal, $u_w$, and vertical, $w_w$, velocity fields from the images.  The PIV software calculated velocities for blocks of 16 pixels $\times$ 16 pixels (i.e., 0.5 cm $\times$ 0.5 cm).  Assuming that the PIV algorithm calculates velocities accurate to 1 pixel per frame, we anticipate a velocity resolution of $\pm 0.9$ cm s$^{-1}$.

\begin{figure}
	\centering{\includegraphics[scale=0.09]{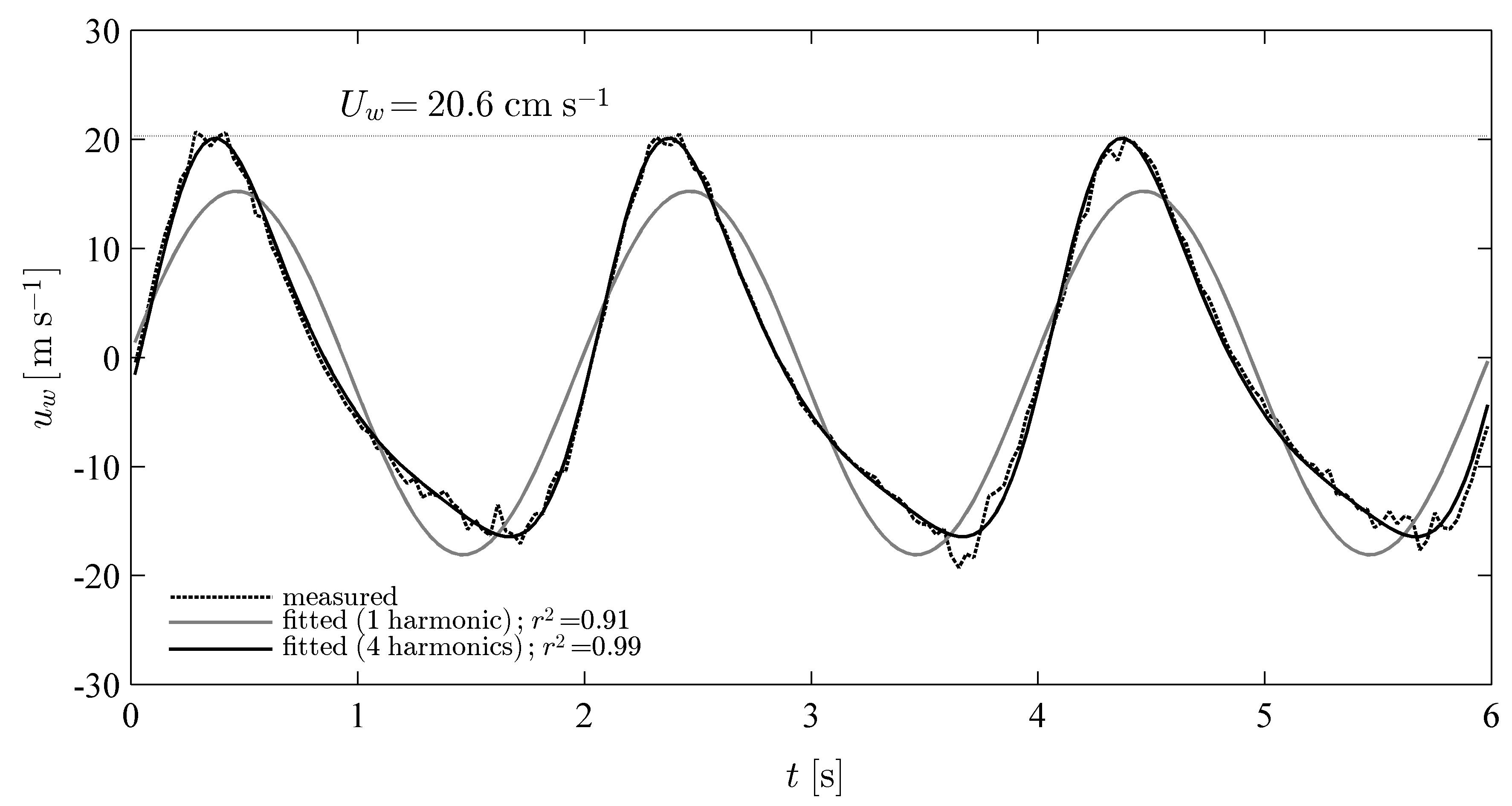}}
	\caption{PIV-measured horizontal wave velocity, $u_w$ (dashed black line), for waves of period $T_w = 2.0$ s and amplitude $a_w\approx 4$ cm.  Also shown is the fitted sinusoidal velocity employing only one harmonic (solid gray line), as well as the fitted velocity employing the first four harmonics (solid black line).  These measurements and fits correspond to the vertical location $z=0$, i.e. the base of the blade.}
	\label{fig:uwfit}
\end{figure}

To characterize the wave-induced flow field, we used the velocities measured approximately 15 cm upstream of the model blade.  At this point, the measured velocities were relatively smooth sinusoidal time series.  Velocities measured closer to the blade were less smooth because of the vorticity generated by the blade itself.  Further, the presence of the blade, blade holder, and wave gage in the field of view led to noisier PIV estimates because, unlike the Pliolite seeding particles, these elements do not track the local flow field.  To estimate the local magnitudes, $U_w$ and $W_w$, of the wave-induced oscillatory velocities, $u_w(z,t)$ and $w_w(z,t)$, we fitted sinusoids to the velocity measurements (Fig.~\ref{fig:uwfit}) at vertical locations ranging from the base of the blade, $z = 0$ cm, to $z = 24$ cm (i.e. 4 cm above the tallest blade length).

For all the wave conditions, we found that the first four harmonics adequately captured the temporal variation in velocity.  The use of higher harmonics did not significantly improve the fits as any further differences between the measured and fitted velocities stemmed from high-frequency turbulent fluctuations or noise (Fig.~\ref{fig:uwfit}).  Importantly, the existence of more than one harmonic in the velocity field indicates that the flow field, though oscillatory, is not symmetrical.  As illustrated in Fig.~\ref{fig:uwfit}, the magnitude of the forward velocity under the wave crest (i.e. positive $u_w$) was generally larger than the magnitude of the backward velocity under the wave trough.  As expected for wave-induced flows, the velocity fields also exhibited some frequency-dependent variability in the vertical direction.  Specifically, the magnitude of the horizontal velocity increased by approximately 6-11\% from $z=0$ to $z = 20$ cm for waves of period $T_w = 2.0$ s, by 19-24\% for waves of period $T_w = 1.4$ s, and by 37-48\% for waves of period $T_w = 1.1$ s.  Similarly, the magnitude of the vertical velocity increased from being near-zero at the base of the blade to roughly 36-40\%, 43-48\%, and 78-83\% of the horizontal velocity at $z=20$ cm in waves of period 2.0 s, 1.4 s, and 1.1 s, respectively.  These frequency-dependent increases are consistent with linear wave theory but do not correspond exactly to theoretical predictions \citep[see e.g.][]{Mei2005}.  However, this lack of agreement is to be expected since the presence of additional harmonics in the flow-field indicates that nonlinear effects are important, and because the waves are likely to be transforming over the trapezoidal box used to house the force sensor and blade assembly (Fig.~\ref{fig:lab}).

The numerical model described in $\S$\ref{sec:theory} was forced with the $z$-dependent sinusoidal fits to the PIV velocity measurements.  For all the comparisons between model predictions and experimental data, the numerical predictions have been phase shifted to account for the 15 cm separation between the PIV velocity measurements and the location of the flexible blade.

\subsection{Dimensionless parameter ranges}\label{sec:dimensionless}
The estimated buoyancy parameter ranged from $B = 0.002$ to $B = 0.15$ for the HDPE blades, and from $B = 2.7$ to $B = 170$ for the silicon foam blades (Table~\ref{tab:lab}).  The dimensionless parameters $Ca$, $KC$ and $L$ were estimated using the maximum fitted horizontal velocity $U_w$ for the vertical location corresponding to the base of the blade (i.e., at $z \approx 0$).  For the wave conditions tested here, the Keulegan-Carpenter number was $KC = 3.7 - 20.6$.  The Cauchy number was $Ca = 0.02 - 36$ for the HDPE blades, and $Ca = 1.0 - 1200$ for the foam blades (see Table~\ref{tab:lab}).  The ratio of blade length to wave excursion, $L = l/A_w$, was smallest for the 5 cm blades in waves of period $T_w = 2.0$ s and amplitude $a_w\approx 4$ cm, with $L = 0.8$.  $L$ was largest for the 20 cm blades in waves of period $T_w = 1.1$ s and amplitude $a_w\approx 2$ cm, with $L = 17$.  

Note that there is a degree of ambiguity in the definition of the above dimensionless parameters due to the vertical variability in the wave-induced flow field.  For example, the 15 cm blades in waves of amplitude $a_w = 4$ cm and period $T_w = 2.0$ s have Cauchy numbers comparable to the 20 cm blades in waves of amplitude $a_w = 4$ cm and period $T_w = 1.1$ s (Table~\ref{tab:lab}).  However, the actual flow fields experienced by the blades (i.e. relative importance of $u_w$ and $w_w$, and the average velocity magnitude over the blade length) are quite different since the higher-frequency waves exhibit much greater variability over depth.  The magnitude of the horizontal velocity increases by $<10\%$ from $z = 0$ to $z = 15$ cm for the $T_w = 2.0$ s waves, but by $\approx 40\%$ from $z=0$ to $z=20$ cm for the $T_w = 1.1$ s waves.  This increase in $u_w$ would result in a substantial increase in drag over the length of the blade, and so the \textit{effective} Cauchy number would be higher for the 20 cm long blades under higher-frequency wave forcing.  In principle, it is possible to employ an average velocity over the blade length to define dimensionless parameters such as the Cauchy number.  However, defining this average velocity scale for flexible blades deforming in a spatially-varying flow would require additional assumptions (e.g. the average height of the blade over a wave cycle).  As a result, we suggest that the magnitude of the horizontal velocity at the base of the blade is a more objective and consistent measure for the velocity scale.

The experiments were designed to have some overlap with conditions found in natural seagrass systems.  Due to variations in material properties, morphology and flow conditions, parameters such as $B$, $Ca$, $KC$ and $L$ vary significantly in the field.  For example, the density of the seagrass \textit{Zostera marina} varies in the range $700-900$ kg m$^{-3}$ \cite{Abdelrhman2007,Fonseca2007}, so that $\rho - \rho_v \approx 100 - 300$ kg m$^{-3}$.  The elastic modulus is estimated to be $E \approx 0.4 - 2.4$ GPa \citep{Bradley2009} and reported blade lengths range from $l \approx 15 - 200$ cm \citep{Ghisalberti2002}.  Limiting the blade length range to $l = 30 - 60$ cm, and assuming the blade width and thickness are $b = 0.8$ cm and $d = 0.35$ mm \citep{Luhar2010}, the buoyancy parameter (Eq.~\ref{eq:B}) is estimated to be $B \approx 1 - 170$.  For a typical velocity range of $U_w = 5 - 100$ cm s$^{-1}$, the Cauchy number (Eq.~\ref{eq:Ca}) ranges from $Ca \approx 10 - 160,000$.  Assuming a peak wave period in the range $T_w \approx 1-8 s$ \citep{Bradley2009,Luhar2013}, the Keulegan-Carpenter number and length ratio are estimated to range between $KC \approx 6-1000$ and $L \approx 0.2-70$. 

Salt marsh vegetation experiences a similarly wide range of hydrodynamic conditions, especially during storms.  However, the plants typically found in salt marshes (e.g. sedges) are characterized by stems of diameter $d \approx 2-3$ mm, roughly an order of magnitude larger than typical seagrass blade thicknesses.  This translates into larger flexural rigidities and lower Cauchy numbers compared to seagrasses \citep{Mullarney2010}. (Note: $I \propto bd^3$ for rectangular cross-sections and $I \propto d^4$ for cylindrical cross-sections, where $d$ is either blade thickness or stem diameter.) 

\subsection{Time scales}\label{sec:timescales} 
Several different time-scales influence the unsteady flow-structure interaction considered in this paper: the wave frequency ($f = 1/T_w$), the natural frequency of the blades ($f_n$), and the vortex shedding frequency ($f_v$).  There is potential for resonance or lock-in phenomena if the natural frequency of the flexible blades matches the wave frequency or the vortex shedding frequency, which could significantly alter blade motion \citep{Blevins1984}.  The undamped natural frequency of cantilevered flexible beams in a fluid is given by:

\begin{equation}\label{eq:natfreq}
f_{n} = C_{n} \sqrt{\frac{EI}{l^4(\rho_v bd + \rho C_M(\pi b^2/4))}},
\end{equation}

{\noindent}where $C_n = 0.56$ is a constant.  Assuming $C_M = 1$, the fundamental natural frequencies range from approximately $f_n = 0.11$ Hz for the 20 cm HDPE blades to $f_n = 1.76$ Hz for the 5 cm HDPE blades, with $f_n = 0.44$ Hz for the 10 cm HDPE blades coming closest to matching one of the forcing wave frequencies.  Similarly, the natural frequencies increase from $f_n = 0.06$ Hz for the 20 cm foam blades to $f_n = 0.92$ Hz for the 5 cm foam blades, with the latter coming closest to matching one of the forcing wave frequencies.  In sum, resonant dynamics are possible for the 10 cm HDPE blades ($f_n = 0.44$ Hz) in waves of frequency $f = 0.5$ Hz and for the 5 cm foam blades ($f_n = 0.92$ Hz) in waves of frequency $f = 0.9$ Hz.  However, it is important to keep in mind that the flexible blades considered in this paper undergo large deformations and are highly damped due to drag, and as such the linearized analyses leading to the above undamped natural frequency estimates may not apply.

The vortex shedding frequency is often expressed in dimensionless terms as the Strouhal number, which can be defined as $St = f_v b/U_w$ for the system considered here.  The Strouhal number is related to the inverse of the Keulegan-Carpenter number.  For flat plates in steady unidirectional flows, the Strouhal number is $St \approx 0.12-0.16$ \citep{Vogel1994}.  Based on this Strouhal number range, the Keulegan carpenter number should be $KC \approx 6-8$ for the wave and vortex frequencies to match, $f \approx f_v$.  However, the above range for $St$ is for steady, unidirectional flows.  For the unsteady flows considered here, the dip in added mass coefficient near $KC = 18$ indicates the conditions in which vortex shedding plays an important dynamic role (Fig.~\ref{fig:CDCM}b, \citep{Keulegan1958}).

\section{Results}\label{sec:results}
\subsection{Forces and motion over a wave cycle}
\begin{figure}
	\centering{\includegraphics[scale=0.41]{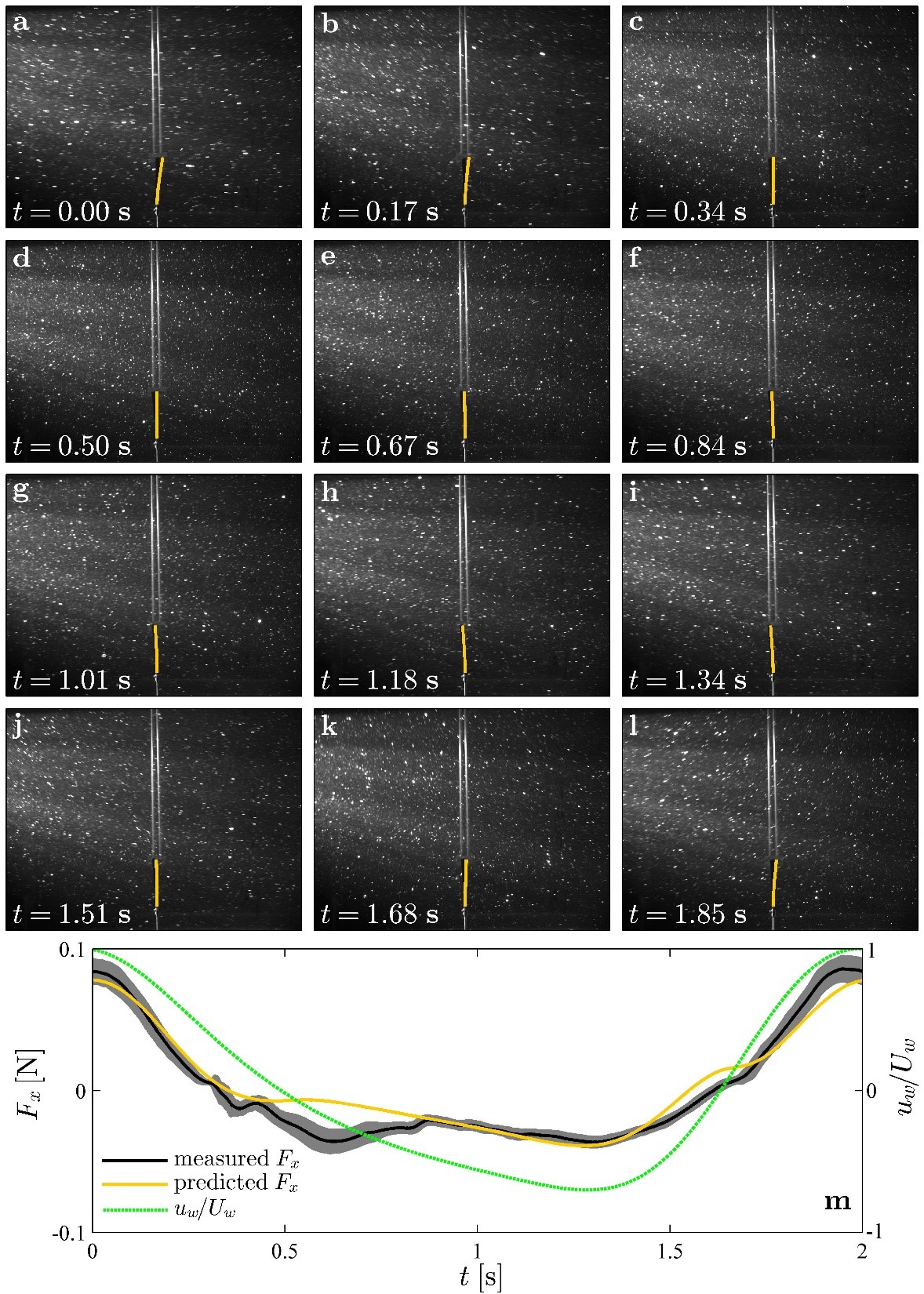}}
	\caption[Blade motion and hydrodynamic force for 5 cm HDPE blade over a wave cycle]{5 cm HDPE blade in waves of period $T_w=2.0$ s and amplitude $a_w\approx 4$ cm.  (a-l) Observed and predicted (yellow) blade posture over the wave-cycle.  Note that the real blade (black) is hidden by the model overlay (yellow) for most of the wave cycle.  The two vertical lines in the background belong to the wave gage used to concurrently measure the wave elevation.  (m) Measured (black) and predicted (yellow) horizontal force, $F_x$, generated by blade.  The shaded gray region represents estimated uncertainty.  The dashed green line shows the normalized horizontal velocity $u_w/U_w$ at the base of the blade.}
	\label{fig:HDPElv05}
\end{figure}

Figure~\ref{fig:HDPElv05} (panels a-l) shows that even for the highest wave velocity tested here ($T_w = 2.0$ s and $a_w\approx 4$ cm, Table~\ref{tab:lab}) the 5 cm HDPE blades did not move significantly in flow.  These observations, also reproduced by the numerical model, are consistent with the behavior expected for low Cauchy numbers ($Ca \le 0.5$ for the 5 cm HDPE blades).  For $Ca<1$, the hydrodynamic forcing is not strong enough to overcome blade stiffness; the blade is essentially rigid.

Importantly, the predicted and measured horizontal forces show good agreement throughout the wave cycle (Fig.~\ref{fig:HDPElv05}m).  This confirms that the Morison force formulation, with values of $C_D$ and $C_M$ based on previous literature, provides a good description of the forces generated by the model blades in oscillatory flows.  There is a $\approx 0.03$ N discrepancy between the measurements and predictions near $t \approx 0.5$ s.  However, such discrepancies are not unexpected given that the Morison force formulation is simply a physically intuitive approximation representing the true time-varying hydrodynamic forces generated by the model blade.  Even for rigid flat plates, the best-fit values of $C_D$ and $C_M$ shown in Fig.~\ref{fig:CDCM} lead to some differences between measured and predicted forces~\cite{Keulegan1958}.  The measured and predicted forces are generally in phase with the velocity, indicating that drag is more important than the inertial forces in this case.

\begin{figure}
	\centering{\includegraphics[scale=0.09]{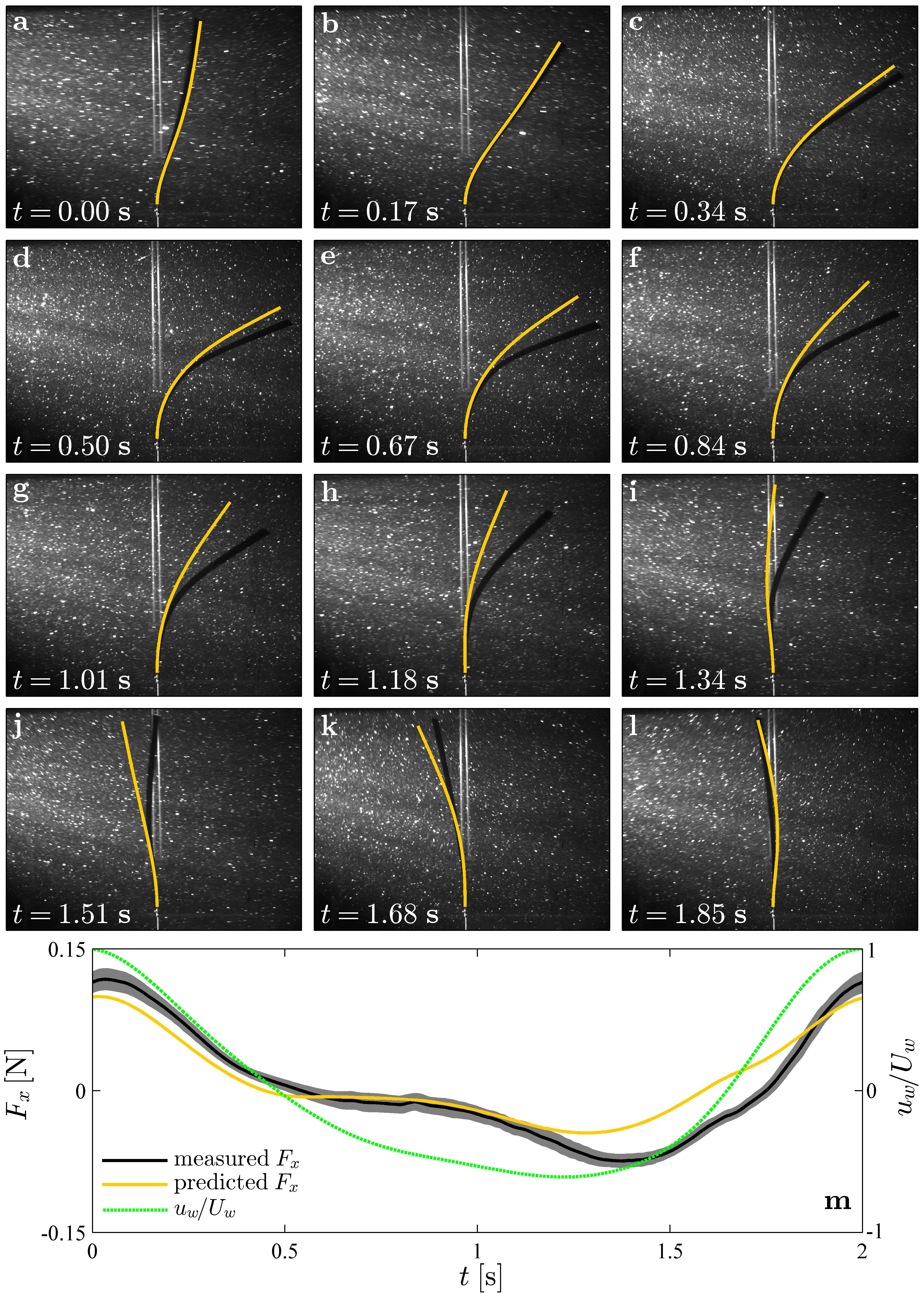}}
	\caption[Blade motion and hydrodynamic force for 20 cm HDPE blade over a wave cycle]{20 cm HDPE blade in waves of period $T_w=2.0$ s and amplitude $a_w\approx 4$ cm.  (a-l) Observed and predicted (yellow) blade posture over the wave-cycle.  (m) Measured (black) and predicted (yellow) horizontal force, $F_x$, generated by blade.  The shaded region represents estimated uncertainty. The dashed green line shows the normalized horizontal velocity $u_w/U_w$ at the base of the blade.}
	\label{fig:HDPElv20}
\end{figure}

\begin{figure}
	\centering{\includegraphics[scale=0.09]{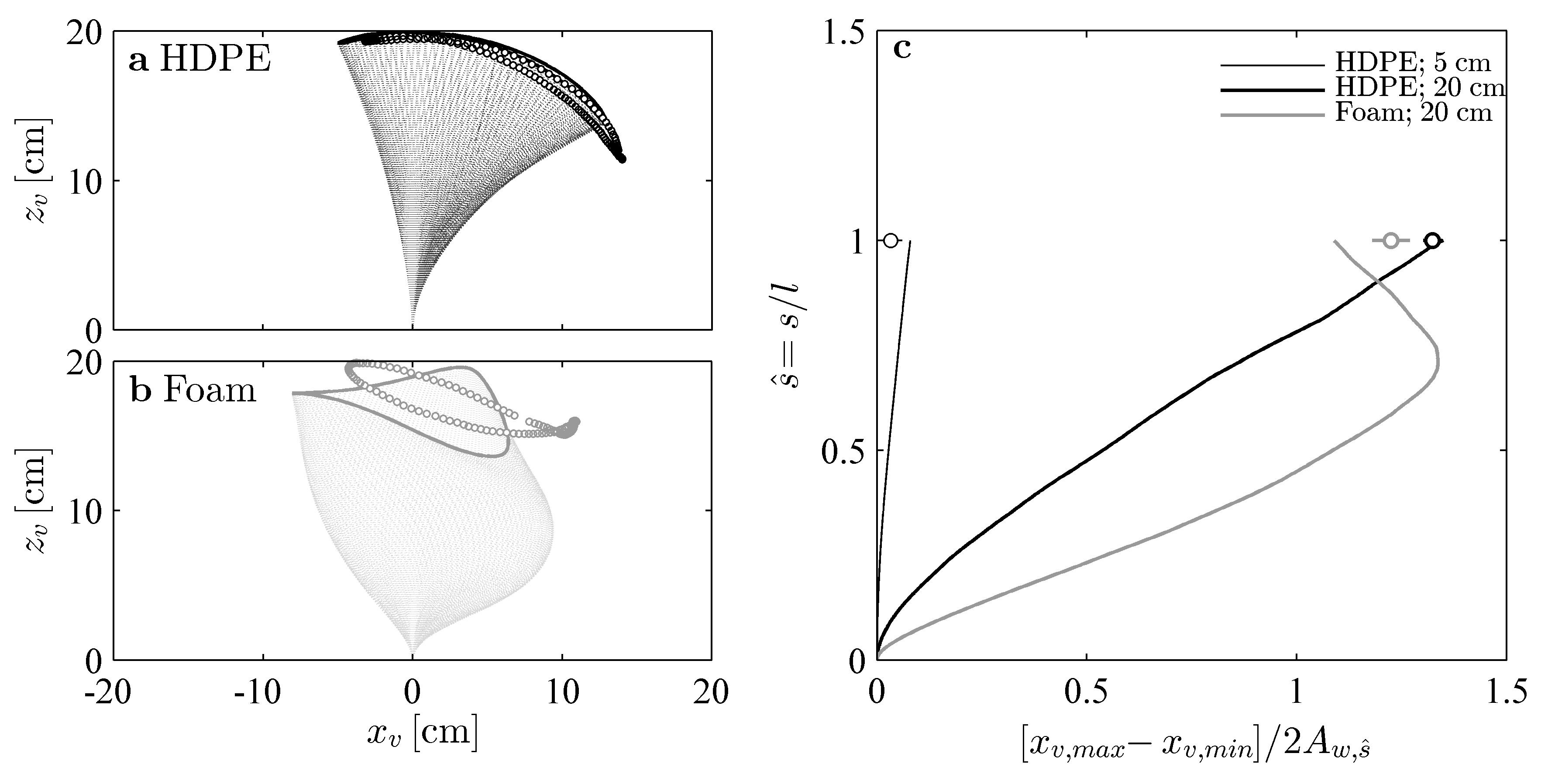}}
	\caption{Blade excursions for the 20 cm long HDPE (a) and Foam (b) blades over a wave cycle.  (c) Profiles of the total horizontal blade excursion over a wave cycle, $[x_{v,max}-x_{v,min}]$, normalized by the \textit{local} wave excursion, $2A_{w,\hat{s}}$, along the blade length, $\hat{s} = s/l$.  Note that $A_{w,\hat{s}}$ was calculated using the measured velocity at the mean vertical position for each $\hat{s}$ over a wave cycle.  In all plots, lines denote outputs from numerical simulations and symbols denote blade tip excursions extracted from the laboratory experiments.  All the data correspond to waves of amplitude $a_w \approx 4$ cm and period $T_w = 2.0$ s.}
	\label{fig:excursion}
\end{figure}

For the same wave condition, the 20 cm HDPE blade moved much more with the wave-induced flow (Fig.~\ref{fig:HDPElv20}, panels a-l) compared to the 5 cm blade.  The total horizontal excursion at the blade tip was $[x_{v,max}-x_{v,min}] \approx 17.5$ cm for the 20 cm blade, while the tip of the 5 cm blade moved back and forth approximately $0.5$ cm.  The wave orbital excursion for these cases ($a_w \approx 4$ cm and $T_w = 2.0$ s) was $2A_w \approx 13$ cm.  So, the tip of the 20 cm HDPE blade moved through a distance roughly 1.3 times the wave excursion, while the tip excursion for the 5 cm HDPE blade was less than $10\%$ of the wave excursion (i.e., the blade remained almost still).  These observations are supported by Fig.~\ref{fig:excursion}c, which shows numerical predictions for the horizontal blade excursion normalized by the local wave orbital excursion for both of these cases.  The predicted excursion for the 5 cm HDPE blade was less than $10\%$ of the wave excursion along the entire blade length, while the excursion for the 20 cm HDPE blade was comparable to the wave excursion for the upper part of the blade (Fig.~\ref{fig:excursion}c, fine and bold black lines).  Specifically, the blade excursion was within $35\%$ of the wave excursion for $\hat{s} > 0.6$, indicating that the upper $40\%$ of the 20cm HDPE blade moved almost passively with the flow.

Passive motion for the upper part of the 20 cm HDPE blade is also reflected in the measured forces.  The maximum measured force for the 20 cm blade was 0.12 N (Fig.~\ref{fig:HDPElv20}m), while the maximum measured force for the 5 cm blade was 0.09 N (Fig.~\ref{fig:HDPElv05}m).  If the 20 cm blade had remained still and upright in the water like the 5 cm blade, we would expect the maximum horizontal force generated to be $4\times 0.09 N \approx 0.36$ N.  $Ca = 36$ for the 20 cm HDPE blade.  At this limit where $Ca \gg 1$, the hydrodynamic forcing is large enough to overcome blade stiffness, and so the upper portion of the blade moves significantly in response to the flow.  Hydrodynamic forces are generated primarily near the base of the blade which remains still relative to the flow.  As noted earlier, this reduction in force can be characterized by the use of an effective rigid blade length $l_e$, which decreases with increasing $Ca$.  We define $l_e$ for wave-conditions in $\S$\ref{sec:le} below, which also considers how $l_e$ varies with the dimensionless parameters $Ca$, $B$ and $L$ in greater detail.

The blade postures and forces predicted by the numerical model show good agreement with the observations for the 20 cm HDPE blade, especially under the wave crest (see $t<0.5$ in Fig.~\ref{fig:HDPElv20}).  For example, the predicted blade tip excursion is 17.8 cm (c.f. the observed 17.5 cm, Fig.~\ref{fig:excursion}a,c), and the maximum predicted force is 0.10 N (c.f. the measured 0.12 N).  Further, the experimental images and numerical results both show that blade posture and motion are asymmetric, such that on average the blade leans in the downstream direction.  We attribute this asymmetric motion primarily to the forward-backward asymmetry in the wave-induced flow field discussed earlier (i.e. magnitude of positive $u_w$ higher than that for negative $u_w$, see also Fig.~\ref{fig:HDPElv20}m).  The asymmetry in blade motion is also reflected in the predicted and measured forces, with both showing a non-zero mean component in the downstream (positive $F_x$) direction.  There are some discrepancies between the measurements and the predictions under the wave trough, $t\approx 1.5$ s in Fig.~\ref{fig:HDPElv20}.  The predicted blade posture is more upright compared to the measurements, and the magnitude of the predicted force is lower by $\approx$0.04 N.  Possible reasons for this discrepancy are discussed in $\S$\ref{sec:discussion}.

\begin{figure}
	\centering{\includegraphics[scale=0.09]{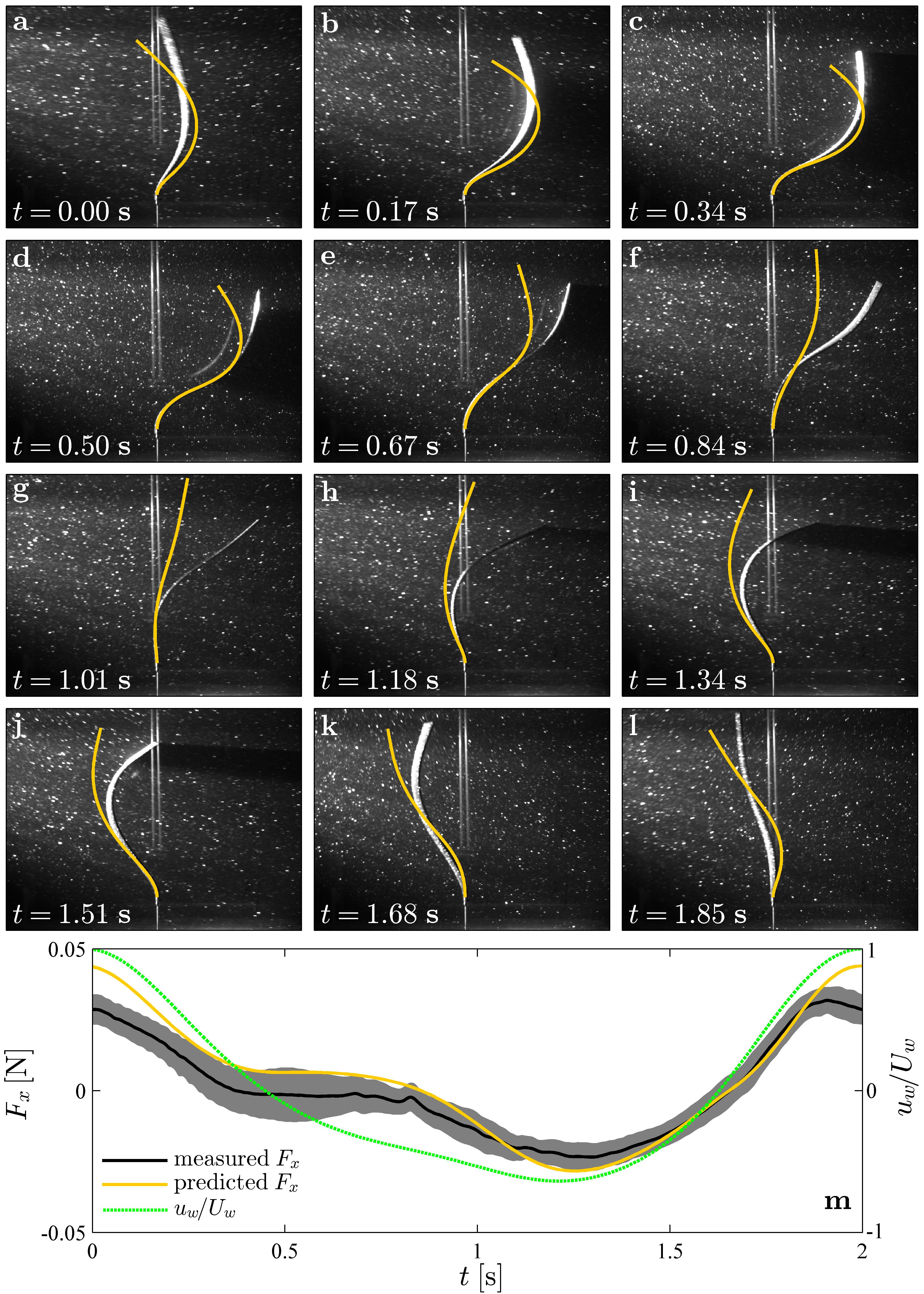}}
	\caption[Blade motion and hydrodynamic force for 20 cm foam blade over a wave cycle]{20 cm foam blade in waves of period $T_w=2.0$ s and amplitude $a_w\approx 4$ cm.  (a-l) Observed and predicted (yellow) blade posture over the wave-cycle.  (m) Measured (black) and predicted (yellow) horizontal force, $F_x$, generated by blade.  The shaded region represents estimated uncertainty. The dashed green line shows the normalized horizontal velocity $u_w/U_w$ at the base of the blade.}
	\label{fig:foamlv20}
\end{figure}

For the wave conditions discussed above ($T_w = 2.0$ s, $a_w \approx 4$ cm), the Cauchy number for the 20 cm foam blade was much higher, $Ca = 1200$.  As a result, a much larger portion of the foam blade (Fig.~\ref{fig:foamlv20}, panels a-l) moved passively with the flow compared to the HDPE blade (see also Fig.~\ref{fig:excursion}b,c).  Recall that the numerically-predicted blade excursion was within $35\%$ of the wave excursion along the upper $\approx 40\%$ of the blade for the 20 cm HDPE blade.  For the 20 cm foam blade, the excursion was within $35\%$ of the wave excursion for $\hat{s}>0.29$ (i.e., along the upper $\approx 70\%$ of the blade, bold gray line in Fig.~\ref{fig:excursion}c).  The force measurements shown in Fig.~\ref{fig:foamlv20}m confirm that a larger portion of the foam blade moves passively in the flow.  The maximum measured force was much lower for the 20 cm foam blade, 0.03 N, compared to the 20 cm HDPE blade, 0.12 N.

For the foam blade, the model predicts blade motion that is more symmetric than the observations.  The simulated blade moves back and forth roughly symmetrically about the vertical, while the real blade leaned to the right near the tip (Fig.~\ref{fig:foamlv20}d,j).   However, the predicted and observed blade excursions are similar over most of the blade.  Because the simulated and real blade move through the same distance over a wave cycle, they experience the same relative velocity.  Since the hydrodynamic force generated by the blade depends on this relative velocity, the predicted (yellow) and measured (black) forces agree within uncertainty through most of the wave cycle (Fig.~\ref{fig:foamlv20}m).

Figures~\ref{fig:excursion}b,c show that the measured blade excursion at the tip of the foam blade (gray circles, $[x_{v,max}-x_{v,min}]/2A_{w,\hat{s}} = 1.23\pm 0.06$) was slightly greater than that predicted by the numerical model (gray line, $[x_{v,max}-x_{v,min}]/2A_{w,\hat{s}} = 1.09$).  We suggest that this discrepancy arises because we do not account for any pressure recovery at the blade tip in the numerical model.  The drag coefficient is assumed to be constant along the entire length of the blade and so the simulated blade experiences greater drag at the tip compared to the real blade.

\subsection{Cycle-averaged forces and blade excursions}
\begin{figure}
	\centering{\includegraphics[scale=0.09]{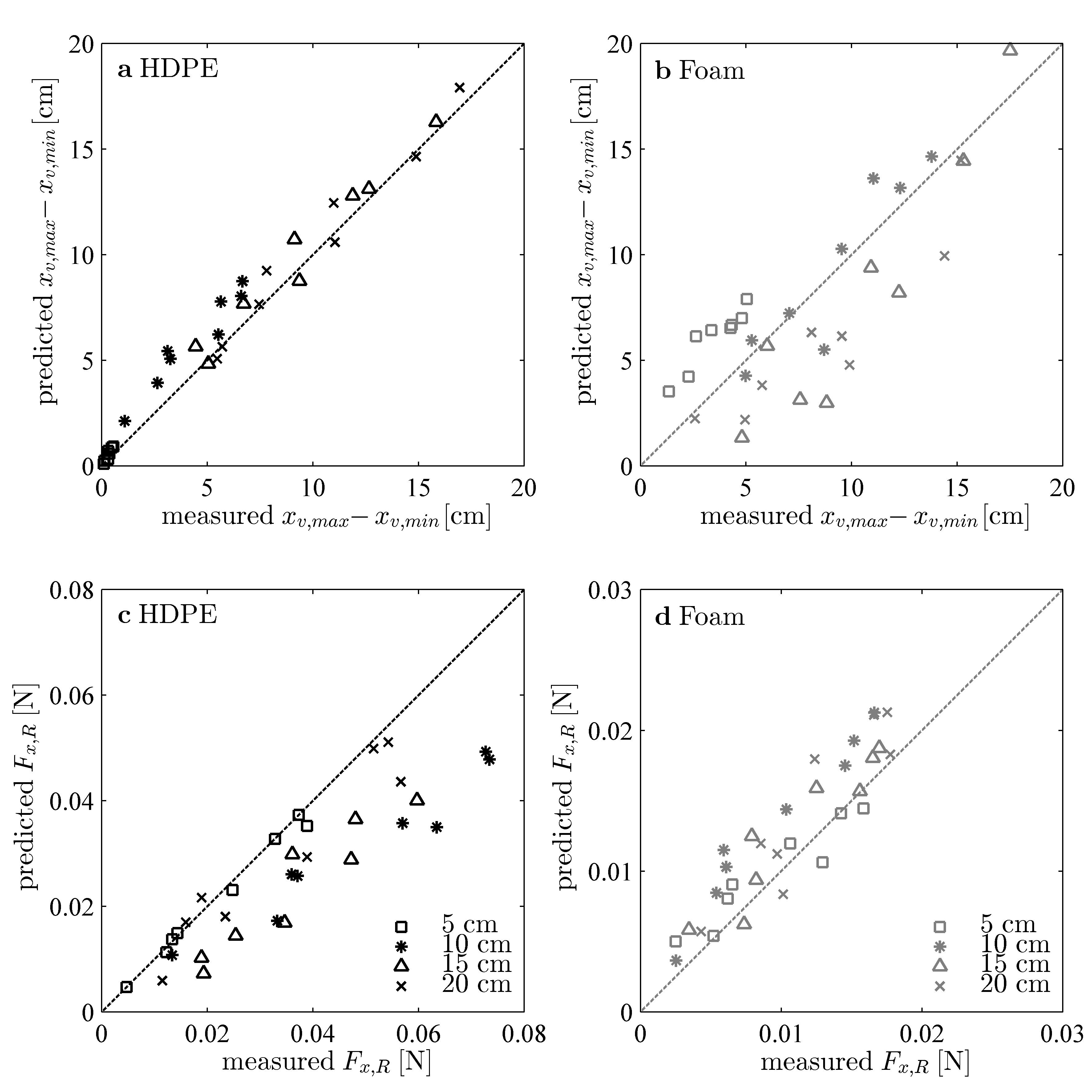}}
	\caption{Predicted horizontal blade-tip excursions $|x_{v,max}-x_{v,min}|$ (a,b) and horizontal forces $F_x$ (c,d) plotted against the measurements from experiments. (a,c) Represent the HDPE blades while (b,d) show results for the Foam blades.  As indicated in the legend, different symbols correspond to different blade lengths, $l$.}
	\label{fig:Fx}
\end{figure}

The three cases described above suggest that the numerical model developed in $\S$\ref{sec:theory} adequately describes the dynamics of flexible blades over the range of conditions tested in the laboratory.  As a further test of model performance, the measured and predicted horizontal excursions at the blade tip $|x_{v,max}-x_{v,min}|$ are compared in Fig.~\ref{fig:Fx}a,b, while the measured and predicted horizontal RMS forces $F_{x,R}$ are compared in Fig.~\ref{fig:Fx}c,d.

Figure~\ref{fig:Fx}a shows that the numerical model does reasonably well in predicting the horizontal excursions of the HDPE blades.  On average, the predicted excursions are larger than the measurements for the shorter 5 cm and 10 cm blades but agree well with the measurements for the longer 15 cm and 20 cm blades.  The ratio of predicted to measured excursions is (mean $\pm$ standard deviation, or s.d.) $1.58 \pm 0.47$, $1.47 \pm 0.29$, $1.08 \pm 0.11$ and $1.03 \pm 0.09$ for the 5 cm, 10 cm, 15 cm and 20 cm blades, respectively.  The large discrepancy and variation in this ratio for the 5 cm HDPE blades is not surprising because the predictions are normalized by the very small measured excursions ($<1$ cm).  The actual difference between the predicted and measured excursions for these short, stiff blades is $0.19\pm 0.17$ cm (mean $\pm$ s.d.), which is comparable to the uncertainty in estimating excursions from the images of blade posture ($\pm 5$ pixels or $\pm 0.2$ cm).

In contrast to the HDPE blades, there are significant differences between the measured and predicted excursions for the foam blades (Fig.~\ref{fig:Fx}b).  Specifically, the numerical model over-predicts blade motion for the short 5 cm blades (gray squares) but under-predicts excursions for the longer 15 cm (triangles) and 20 cm (crosses) blades.  The ratio of predicted to measured excursions is (mean $\pm$ s.d.) $1.85 \pm 0.42$, $1.01 \pm 0.19$, $0.70\pm 0.32$ and $0.69 \pm 0.18$ for the 5 cm, 10 cm, 15 cm, and 20 cm foam blades, respectively.  Again, these discrepancies are thought to arise because the numerical model does not adequately account for edge-effects near the blade tips.  This is illustrated by the discussion presented in \S\ref{sec:modelPerformance}.

Figures~\ref{fig:Fx}c,d show that the model generally under-predicts the forces acting on the HDPE blades and over-predicts forces for the foam blades.  Specifically, the ratio of predicted to measured $F_{x,R}$ was $1.27 \pm 0.30$ (mean $\pm$ s.d.) for the foam blades.  For the HDPE blades, the ratio was $0.78 \pm 0.19$.  Given the 0.001 N resolution and 10$\%$ accuracy of the load cell, it can be argued that the measurements agree with the predictions within uncertainty.  Uncertainty in material properties offers another possible explanation for these discrepancies.  This is especially true for the foam blades, where the uncertainty in both the density difference, $\Delta\rho$, and the elastic modulus, $E$, was greater than 10$\%$ (Table~\ref{tab:lab}).

Upon closer inspection, Fig.~\ref{fig:Fx}c suggests that while the measured and predicted forces agree very well for the 5 cm (squares) and 20 cm (crosses) HDPE blades, the measurements are consistently larger than the predictions for the 10 cm and 15 cm blades.  For example, over all eight wave conditions, the ratio of predicted to measured RMS force was $0.66 \pm 0.09$ (mean $\pm$ s.d.) for the 10 cm blades.  In contrast, this ratio of predicted to measured force was $0.98 \pm 0.05$ for the 5 cm blades and $0.88 \pm 0.20$ for the 20 cm blades.  As discussed below, the 10 cm HDPE blades generate forces larger than those expected even for rigid blades.  This unintuitive behavior is associated with a vortex shedding event, which causes the blade to \textit{spring} backwards (\S\ref{sec:enhancedFx}).  Unfortunately, the numerical model in its present form cannot reproduce this unsteady effect.  This is because the numerical model predicts the blade motion in response to a precribed wave flow field.  It does not capture the impact of the blade on the flow, i.e. it does not recreate the vortex shedding event and subsequent blade dynamics.

\subsection{Effective blade length}\label{sec:le}
For unidirectional flows, the effective length $l_e$ was defined as the length of a rigid, upright blade that generates the same horizontal drag as the flexible blade of length $l$ \citep{Luhar2011}.  Because of the time-varying nature of the hydrodynamic forces generated, the effective length can be defined in multiple different ways for oscillatory flows.  We define the effective length based on the RMS force:

\begin{equation}\label{eq:le}
\frac{l_{e,R}}{l} = \frac{measured\;F_{x,R}}{rigid\;F_{x,R}},
\end{equation}

{\noindent}where the rigid-blade force was calculated using the PIV-measured velocity field $u_w(z,t)$, as:

\begin{equation}\label{eq:Fxrig}
F_x(t) = \int\limits_{0}^{l} \frac{1}{2}\rho C_D b |u_w|u_w + \rho \left(\frac{\pi}{4}b^2C_M + bd \right)\frac{\partial u_w}{\partial t} \, dz.
\end{equation} 

{\noindent}The first term inside the integral is the drag force (Eq.~\ref{eq:fd}), and the second term represents the added mass (Eq.~\ref{eq:fam}) and virtual buoyancy forces (Eq.~\ref{eq:fvb}).  However, one may also define the effective length using, for example, the maximum force over a wave cycle.

Recall from \S \ref{sec:scaling} that once the hydrodynamic forcing exceeds the restoring forces due to blade stiffness and buoyancy (i.e. $Ca \gg 1$ and $Ca \gg B$), we expect that the effective length scales as $l_e/l \sim Ca^{-1/3}$ for the quasi-steady large excursion limit ($L \ll 1$) and $l_e/l \sim (Ca L)^{-1/4}$ for the small-deflection, small excursion limit ($L \gg 1$).  For the laboratory experiments described in $\S$\ref{sec:experiments}, the ratio of blade length to wave excursion ranged from $L = 0.8$ to $L = 17$.

\begin{figure}
	\centering{\includegraphics[scale=0.09]{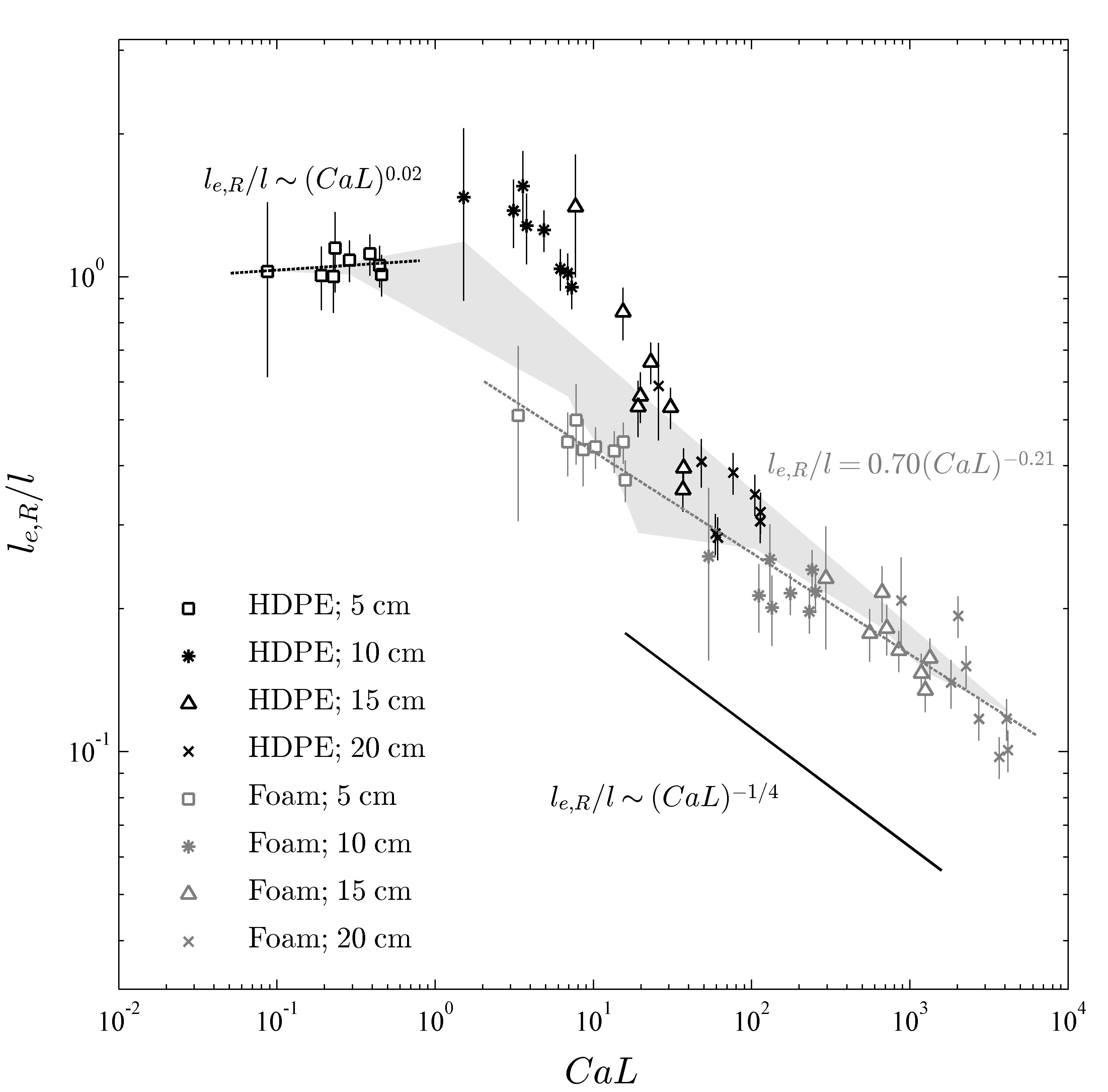}}
	\caption[Effective length plotted against $CaL$]{$l_{e,R}/l$ plotted against $CaL$.  The solid black line shows the expected scaling for small wave ($L\gg 1$) excursions, $l_{e,R}/l\sim (CaL^{-1/4})$.  The dotted gray line shows the best fit to all the foam data.  The dotted black line shows the best fit to the 5 cm HDPE blade data.  The error bars represent the measurement uncertainty (i.e. the 0.001 N resolution and 10$\%$ accuracy, whichever is largest).  Symbols and color scheme as indicated on plot.  The shaded region in the background represents the range of effective lengths predicted by the numerical model.  Individual predictions have not been shown to maintain figure clarity.}
	\label{fig:le}
\end{figure}

The measured effective lengths (Eq.~\ref{eq:le}) for all sixty-four laboratory tests are shown in Fig.~\ref{fig:le}.  First consider the foam blades, for which $Ca \ge 1$ (gray symbols in Fig.~\ref{fig:le}).  The measured effective lengths for these blades collapse together, with a best fit power-law $l_{e,R}/l = 0.70\pm 0.05 (Ca L)^{-0.21\pm 0.02}$, suggesting that the small-excursion scaling $l_{e,R}/l \sim (Ca L)^{-1/4}$ applies even for $L \sim O(1)$.  Of course, this collapse is not perfect.  However, any deviations could easily be attributed to the fact that the scaling law (Eq. \ref{eq:lescaling}) neglects the effects of varying inertial and buoyancy forces across each test case, and that the estimator for the rigid blade drag reference (Eq.~\ref{eq:Fxrig}) does not account for the variation in $C_D$ in time or along the blade length.

For $Ca L < 1$, the blades are essentially rigid in the flow.  This is illustrated with the 5 cm HDPE blades (black squares in Fig.~\ref{fig:le}), for which $l_{e,R}/l$ is approximately constant and equal to 1.  The best-fit power law is $l_{e,R}/l = 1.08 \pm 0.06 (Ca L)^{0.02 \pm 0.04}$ (dashed line in Fig.~\ref{fig:le}), confirming that there is no dependence between $l_{e,R}/l$ and $Ca L$, i.e. the blade behaves like a rigid flat plate.

The 10 cm HDPE blades (black stars in Fig.~\ref{fig:le}) do not conform to the predicted scaling law for either the rigid limit ($l_{e,R}/l \approx 1$) or the small-excursion limit ($l_{e,R}/l \sim (Ca L)^{-1/4}$).  For these cases, the measured effective lengths are larger than unity, indicating that the flexible blades generate \textit{greater} RMS forces than rigid upright blades of similar dimensions.  Observations of blade motion suggest that the magnitude of the hydrodynamic forces generated by the 10 cm blades was enhanced as they sprung backwards in flow following a vortex shedding event (see \S\ref{sec:enhancedFx}).  This may be a consequence of the fact that the drag forcing and stiffness were comparable for these blades, $Ca \sim O(1)$, allowing for more complex time-varying interactions.  The measured effective lengths for the 20 cm HDPE blades (black crosses in Fig.~\ref{fig:le}), for which $Ca \ge 2$, are merging towards the foam blades (gray symbols), suggesting that the scaling $l_{e,R}/l \sim (Ca L)^{-1/4}$ may apply at values of $Ca$ as low as this.

The observed trends do not change significantly if we use the ratio of maximum forces $l_{e,M}/l$ instead of the RMS forces.  Specifically, best-fits are $l_{e,M}/l = 1.05 \pm 0.12 (CaL)^{-0.03 \pm 0.08}$ for the 5 cm HDPE blades, and $l_{e,M}/l = 0.65 \pm 0.07  (Ca L)^{-0.22 \pm 0.02}$ for all the foam blades.  In summary, the results presented in this section indicate that the effective length framework provides a useful method to account for plant motion when predicting the hydrodynamic forces generated by flexible vegetation.

\section{Discussion}\label{sec:discussion}
\subsection{Numerical model performance}\label{sec:modelPerformance}
Without the use of any tuned parameters (recall that $C_D$ and $C_M$ were based on previous literature for flat plates), the numerical model developed in $\S$\ref{sec:theory} predicts the forces generated by the model blades in laboratory tests reasonably well across a range of blade properties and flow conditions.  The ratio of predicted to measured RMS forces for all 64 test cases was distributed with mean and standard deviation $1.03 \pm 0.35$.  This confirms that our model captures the salient physics governing the wave-induced dynamics of flexible blades.  Specifically, our results suggest that rigid-body $C_D$ and $C_M$ may be used for flexible bodies as long as the relative, body-normal velocity and acceleration are used to calculate the drag and added mass forces.  However, it is important to keep in mind that, although it has been used with relative success in previous studies and in this paper, the Morison force formulation is a physically intuitive, simplified representation of the true unsteady forces acting on the blade.  There are often large differences between the true force generated by the body and predictions made by summing the drag and added-mass terms \citep{Keulegan1958} with constant $C_D$ and $C_M$ (see e.g. Fig.~\ref{fig:HDPElv05}m).

\begin{figure}
	\centering{\includegraphics[scale=0.09]{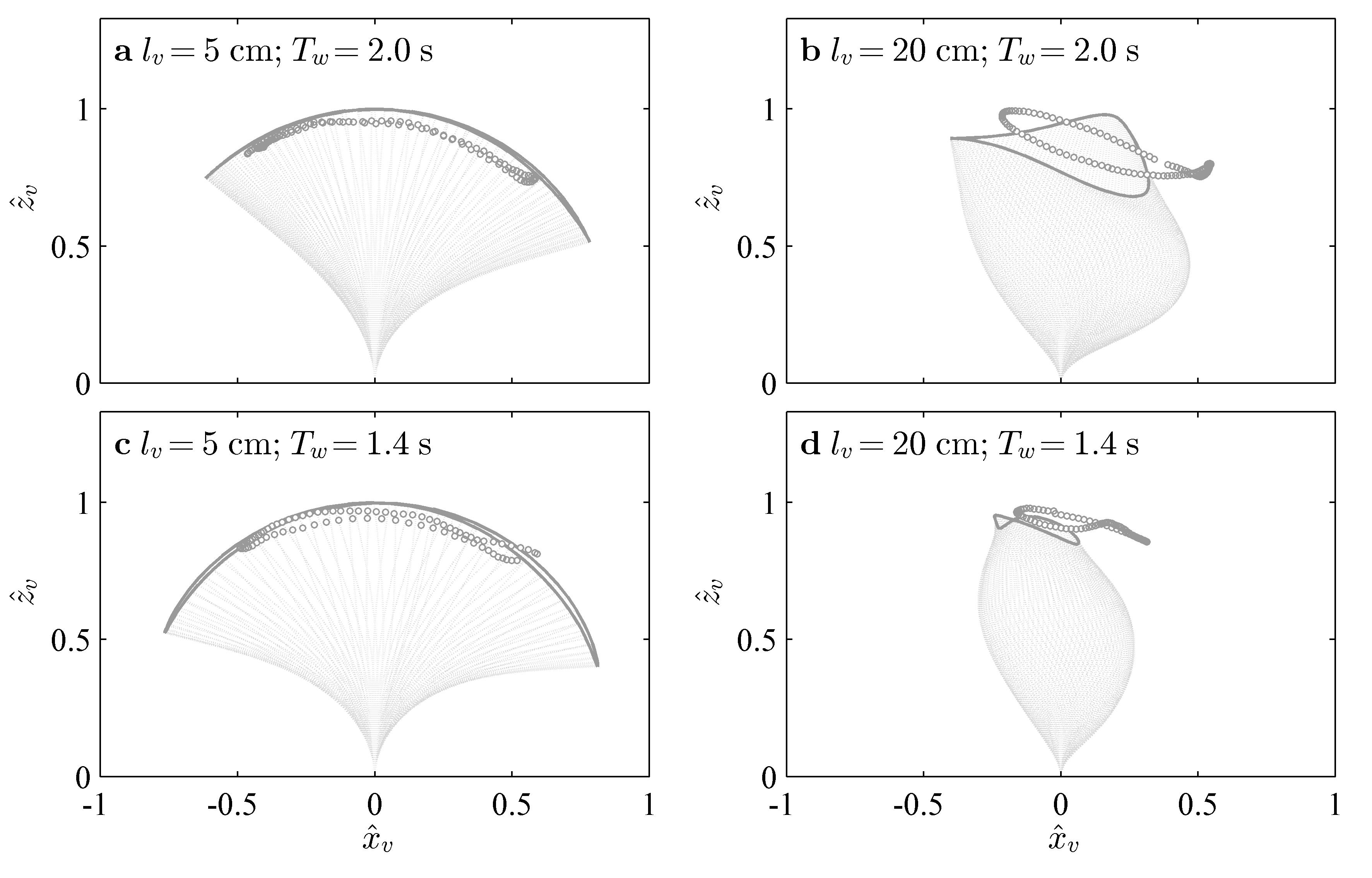}}
	\caption{Normalized blade excursions for the 5 cm (a,c) and 20 cm (b,d) long foam blades over a wave cycle.  Lines denote outputs from numerical simulations and symbols denote blade tip excursions extracted from the laboratory experiments.  All the plots correspond to waves of amplitude $a_w \approx 4$ cm.  Plots (a,b) show results corresponding to waves of period $T_w = 2.0$ s, while (c,d) show results for $T_w=1.4$ s.}
	\label{fig:foamExcursions}
\end{figure}

The previous section showed that there are significant differences between the predicted and observed blade postures for the highly flexible foam blades, especially near the blade tip (see Fig.~\ref{fig:foamlv20} and Fig.~\ref{fig:excursion}b).  Specifically, the excursion of the blade tips is over-predicted by the numerical model for the shorter 5 cm foam blades but under-predicted for the 15 cm and 20 cm blades (Fig.~\ref{fig:Fx}b).  We suggest that these discrepancies may arise because we assume that $C_D$ and $C_M$ are constant over the length of the blade.  In reality, pressure recovery near the blade tip leads to a reduction in forces, and therefore a reduction in the effective $C_D$ and $C_M$.  In addition, observations of blade posture indicate that the longer foam blades tend to twist near the tip.  This twisting, which cannot be captured in the present two-dimensional model, could lead to a further reduction in the hydrodynamic forces via a reduction in frontal area.  Figure~\ref{fig:foamExcursions} illustrates why larger modeled forces at the blade tip result in the discrepancies between the predicted and measured excursions.  The excursion of the shorter 5 cm blades is largest at the blade tip (Fig.~\ref{fig:foamExcursions}a,c).  In this case, larger forces at the tip translate directly into greater blade deflection.  This results in predicted blade tip excursions that are larger than the measurements.  In contrast, the horizontal excursion peaks somewhere along the mid-span for the longer 20 cm foam blades (Fig.~\ref{fig:foamExcursions}b,d).  Here, larger forces near the tip serve to inhibit motion, resulting in predicted excursions that are smaller than the measurements.

Finally, we note for completeness that the numerical model developed in this paper does not account for the hydrodynamic force generated due to blade curvature, often termed the \textit{reactive} force \citep{Buchak2010}.  This force arises as the component of flow parallel to the blade accelerates to follow the blade shape.

\subsection{Flexibility can enhance forces}\label{sec:enhancedFx}

\begin{figure}
	\centering{\includegraphics[scale=0.09]{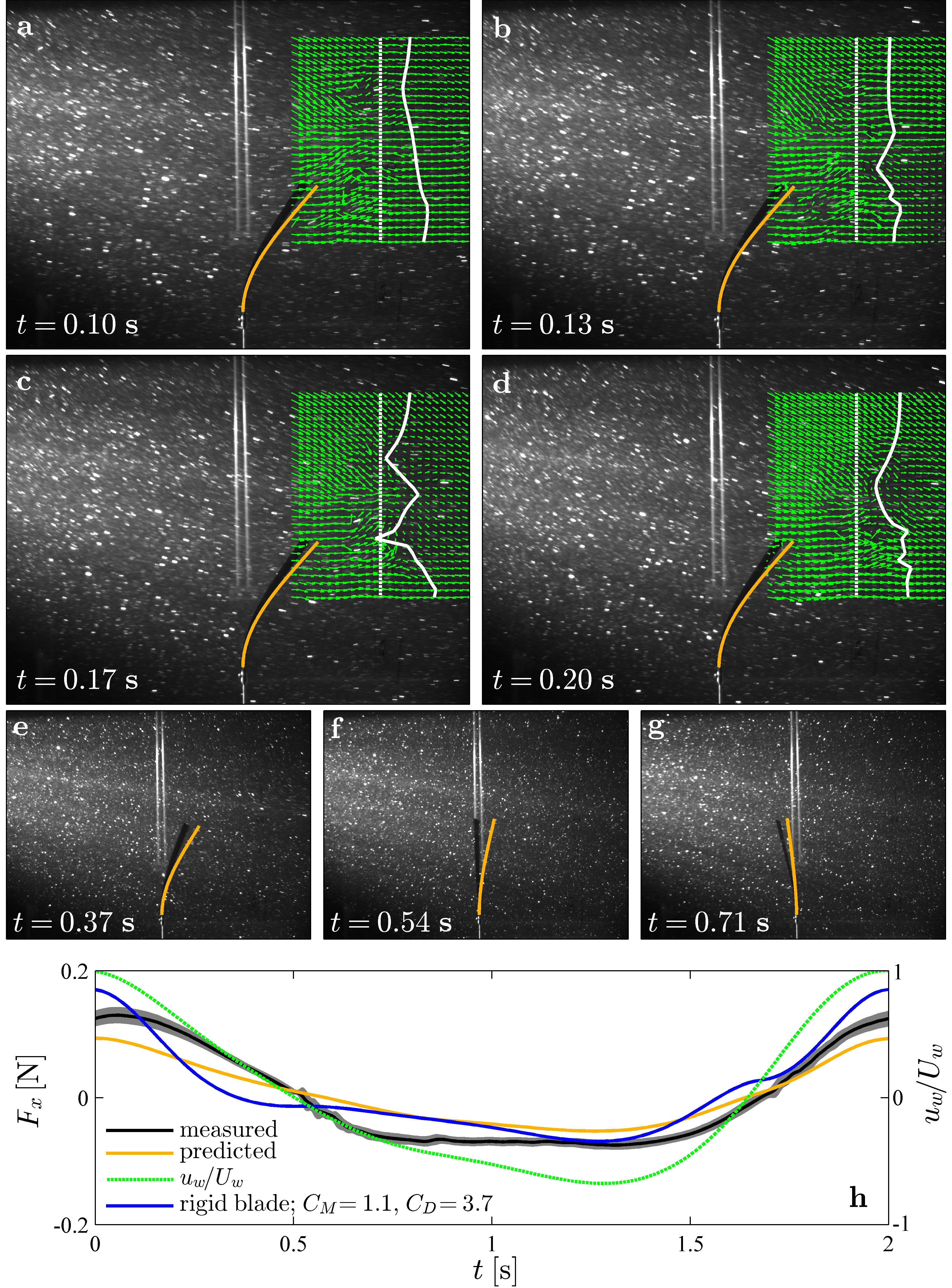}}
	\caption[Vortex shedding from 10 cm HDPE blade]{(a-d) Blade postures and PIV velocity field (green arrows) for 10 cm HDPE blade between $t = 0.1 - 0.2$ s in waves of period $T_w = 2.0$ s and amplitude $a_w\approx 4$ cm.  The solid white line shows the variation in horizontal velocity along the dotted white line.  (e-g) Vortex shedding leads to the HDPE blade springing backwards between $t \approx 0.3 - 0.6$ s.  (h) A comparison of measured (black), predicted (yellow), and rigid-blade (blue) forces. The dashed green line shows the normalized horizontal velocity $u_w/U_w$ at the base of the blade.}
	\label{fig:HDPElv10}
\end{figure}

Recall that the measured forces were significantly larger than the predictions for the 10 cm HDPE blades (Fig.~\ref{fig:Fx}a), such that the ratio of predicted to measured $F_{x,R}$ was distributed with mean and standard deviation $0.66 \pm 0.09$.  For these blades, the Cauchy number ranged from $Ca = 0.24$ to $Ca = 4.0$ (Table~\ref{tab:lab}).  For $Ca\sim O(1)$, the hydrodynamic forcing and the restoring force due to blade stiffness are comparable.  As a result, there can be a transition between forcing-dominated and stiffness-dominated conditions within a wave cycle.  As illustrated by Fig.~\ref{fig:HDPElv10}, this transition leads to unsteady blade behavior and an enhancement of forces that cannot be reproduced by the numerical model.  Specifically, a vortex is shed from the blade between time $t=0.10 - 0.20$ s (Fig.~\ref{fig:HDPElv10}a-d, see sharp velocity gradients and evidence of flow reversal in the PIV measurements).  Immediately after this, between $t = 0.3 - 0.6$ s (Fig.~\ref{fig:HDPElv10}e-g), the real blade (shown in black) \textit{springs} backwards more rapidly than the modeled blade (shown in yellow).  As the blade springs backwards, there is significant relative motion between the blade and the water, and this leads to the generation of additional hydrodynamic drag (Fig.~\ref{fig:HDPElv10}h).  Specifically, between $t=0.3$ s and $t = 0.6$ s, the measured drag is elevated compared to the model and rigid blade drag.  In contrast to the real blade, the simulated blade moves back gradually in the flow (yellow lines in Fig.~\ref{fig:HDPElv10}e-g).  Hence, the relative motion between the blade and the water is lower for the simulations, and so is the predicted force (yellow line in Fig.~\ref{fig:HDPElv10}h).

We suggest that the numerical model cannot reproduce the observed behavior because it employs constant $C_D$ and $C_M$.  In effect, the shedding event may lead to a local (in time) reduction in the added mass.  Recall that the rigid flat plate data also show a reduction in $C_M$ for the conditions in which a single eddy is shed from the plate in each wave half-cycle (Fig.~\ref{fig:CDCM}b, $KC \approx 18$).  To test this hypothesis, we fitted drag and added mass coefficients to the force and velocity measurements shown in Fig.~\ref{fig:HDPElv10}h, assuming that the blade remains rigid and upright, and that $u_w$ does not vary over the blade length.  In other words, we identified the combination of $C_D$ and $C_M$ that provides the best fit to the measured $F_x(t)$ under the assumption:

\begin{equation}\label{eq:FxFit}
F_x(t) = \frac{1}{2}\rho C_D b l |u_w|u_w + \rho \left(\frac{\pi}{4}b^2 l C_M + b d l \right)\frac{\partial u_w}{\partial t},
\end{equation}

{\noindent}where $u_w(t)$ corresponds to the velocity at the base of the blade, $z=0$.  The best-fit values were $C_D = 3.9$ and $C_M = -0.1$.  For reference, $C_D = 3.7$ and $C_M = 1.1$ for a rigid flat plate under similar wave conditions ($KC \approx 20$, see Fig.~\ref{fig:CDCM}).  The negative fitted value for $C_M$ supports the hypothesis that vortex shedding event modifies the inertial forces acting on the blade (and perhaps even imparts momentum to the blade).  Consistent with a negative $C_M$, Fig.~\ref{fig:HDPElv10}h shows that the peak in the measured $F_x$ (black line) lags behind the peak in $u_w$ (green line) by $\approx 0.1$ s, which coincides with a period of negative $\partial u_w/\partial t$.

Similar shedding events were also observed for the 5 cm and 20 cm HDPE blades.  However, because $Ca < 1$ for the 5 cm blade, the vortex shedding did not impact the blade, which remained motionless throughout the wave cycle, i.e., the blade was essentially rigid.  For the 20 cm blade, the Cauchy number was $Ca \gg 1$, and so we suggest that this blade was not stiff enough to spring backwards following the shedding event.  For the same wave condition, the numerical model was able to predict the measured forces reasonably well for both the 5 cm (Fig.~\ref{fig:HDPElv05}) and 20 cm (Fig.~\ref{fig:HDPElv20}) HDPE blades.  Given that the natural frequency of the 10 cm HDPE blades, $f_n = 0.44$ Hz, was close to the forcing frequency of the $T_w = 2.0$ s waves, resonance might offer another possible explanation for the observed unsteady behavior and enhancement of forces.  However, the ratio of predicted to measured $F_{x,R}$ is not significantly different for the $T_w = 2.0$ s wave conditions compared to the remaining cases, for which resonance is not expected ($0.70 \pm 0.08$ vs. $0.61 \pm 0.09$), which suggests that resonance is not important.

Interestingly, because of the unsteady blade behavior described in this section, the RMS force generated by the flexible 10 cm HDPE blade was greater than that predicted for a rigid 10 cm blade (blue line in Fig.~\ref{fig:HDPElv10}h).  Specifically, $F_{x,R}=0.074$ N for the flexible blade, while we expect $F_{x,R}=0.070$ N for a rigid blade.  Indeed, for most of the laboratory tests with the 10 cm HDPE blades, all of which have $Ca\sim O(1)$, the measured $F_{x,R}$ was greater than that predicted for a rigid blade (black stars in Fig.~\ref{fig:le}), which we suggest is associated with the blade response to vortex shedding described above and illustrated in Fig.~\ref{fig:HDPElv10}.

\subsection{Buoyancy effects}\label{sec:buoyancy}
\Citet{Luhar2011} showed that blade buoyancy delayed the onset of reconfiguration in steady flows until the Cauchy number exceeded the value of the buoyancy parameter, $Ca > B$.  However, buoyancy does not seem to play as important a role for the wave-induced oscillatory flows considered here.  Specifically, the buoyancy parameter ranged from $B = 2.7 - 170$ for the foam blades, with $Ca < B$ in some cases (Table~\ref{tab:lab}).  Yet, the measured effective lengths all collapse onto a single line (Fig.~\ref{fig:le}).  These observations can be explained by the fact that, for wave-induced oscillatory flows with $L > 1$, the blades remain relatively upright in the flow.  For upright blades the contribution of buoyancy to the blade-normal force balance, which is primarily responsible for dictating blade motion, is negligible.  In the governing equation, Eq.~\ref{eq:governdim}, the buoyancy term is: $iBe^{i\theta}$.  The blade-normal (i.e., real) component of this term is $-B\sin\theta$.  As in \S\ref{sec:scaling}, for $L > 1$ we expect that the angle $\theta \sim L^{-1} \ll 1$, and so $B\sin\theta \sim B\theta \sim (B/L)$.  At this limit therefore, buoyancy is only important as long as $(B/L) > Ca$, or $(CaL) < B$.  While $Ca < B$ for some of the cases tested in the laboratory, $CaL > B$ for all the cases.  Therefore, buoyancy should not play a significant role, consistent with the observations.

\section{Conclusion}\label{sec:conclusion}
Given the ability of aquatic plants to stabilize the substrate, filter nutrients from the water column and provide habitat, recent studies suggest that the creation and restoration of vegetated ecosystems may provide cost-effective, sustainable, and ecologically sound coastal protection in the face of rising sea-levels and climate change \citep{Temmerman2013}.  A significant component of the coastal protection provided by vegetated ecosystems arises from their ability dissipate wave energy by generating hydrodynamic drag.  However, significant challenges remain in developing accurate wave dissipation models for such systems.

In particular, as noted earlier, there is no consistent framework to account for plant flexibility and motion.  Without this framework, it is difficult to make comparisons across species and systems, which means that most studies are limited to using drag coefficients fitted to measurements.  Such drag coefficient calibrations typically employ the Reynolds number, $Re$, or Keulegan-Carpenter number, $KC$, as the independent governing parameters \citep{Kobayashi1993,Bradley2009, Mendez2004}.  $Re$ and $KC$ can account for the variation in drag with hydrodynamic conditions (see e.g., Fig~\ref{fig:CDCM}).  However, they cannot account for any drag reduction due to plant flexibility because they do not reflect the underlying physics.  Given the likely variation in vegetation stiffness and buoyancy, the calibrated drag coefficient for one species will not hold for another species.  Instead, we suggest the use of the effective length concept proposed by \citet{Luhar2011} and developed further in this paper to account for vegetation motion.  The effective length approximates the length of blade over which relative motion between the blades and the water is significant.  So wave energy dissipation within the meadow can be calculated by assuming that the vegetation is rigid, but of length $l_e$, rather than $l$.

Importantly, a characterization of $l_e/l$ as a function of the dimensionless parameters that govern blade motion $Ca$, $B$, and $L$, is likely to hold across systems.  A reanalysis of existing wave decay datasets that translates the fitted drag coefficients into effective lengths, and considers the variation of these effective lengths with parameters such as $Ca$, $B$ and $L$ would be a useful first step towards providing a consistent methodology for future work.  For field studies, $Ca$, $B$, and $L$, may be calculated based on measured vegetation properties, and the significant wave height, $H_S$, and peak period, $T_P$.  Note that defining an effective length can become complicated for conditions in which there is significant variation in the wave-induced flow field over depth (e.g. for marsh environments with short fetches and high-frequency waves).  As discussed in \S\ref{sec:dimensionless}, there is some ambiguity in the definition of the velocity scale for depth-varying flows.  Further, the effective length may be different depending on whether it is used to quantify a reduction in drag ($\propto u_w^2$) or wave energy dissipation ($\propto u_w^3$).  The broadband nature of waves in the field also poses additional challenges.  As observed by \citet{Bradley2009}, the blades may move in response to secondary frequencies rather than the peak frequency.  For such cases, the motion of the blades is not in phase with the water, resulting in some relative motion over the entire blade length even if $l_e < l$.  However, this effect may be accounted for by employing a frequency-dependent effective length.

\section*{Acknowledgments}
Financial support from the National Science Foundation under grant OCE 0751358 is gratefully acknowledged.  The authors are also grateful for feedback from three anonymous reviewers that helped improve this manuscript.

\bibliographystyle{elsarticle-harv}
\bibliography{2015-Luhar-WaveDynamics-Rev2}

\begin{thebibliography}{38}
\expandafter\ifx\csname natexlab\endcsname\relax\def\natexlab#1{#1}\fi
\expandafter\ifx\csname url\endcsname\relax
  \def\url#1{\texttt{#1}}\fi
\expandafter\ifx\csname urlprefix\endcsname\relax\def\urlprefix{URL }\fi

\bibitem[{Abdelrhman(2007)}]{Abdelrhman2007}
Abdelrhman, M.~A., 2007. Modeling coupling between eelgrass zostera marina and
  water flow. Marine Ecology-Progress Series 338, 81--96.

\bibitem[{Alben et~al.(2002)Alben, Shelley, and Zhang}]{Alben2002}
Alben, S., Shelley, M., Zhang, J., 2002. Drag reduction through self-similar
  bending of a flexible body. Nature 420~(6915), 479--481.

\bibitem[{Batchelor(2000)}]{Batchelor2000}
Batchelor, G.~K., 2000. An introduction to fluid dynamics. Cambridge University
  Press.

\bibitem[{Blevins(1984)}]{Blevins1984}
Blevins, R.~D., 1984. Applied fluid dynamics handbook.

\bibitem[{Bradley and Houser(2009)}]{Bradley2009}
Bradley, K., Houser, C., 2009. Relative velocity of seagrass blades:
  Implications for wave attenuation in low-energy environments. Journal of
  Geophysical Research-Earth Surface 114, F01004.

\bibitem[{Buchak et~al.(2010)Buchak, Eloy, and Reis}]{Buchak2010}
Buchak, P., Eloy, C., Reis, P.~M., 2010. The clapping book: Wind-driven
  oscillations in a stack of elastic sheets. Physical Review Letters 105,
  194301.

\bibitem[{Costanza et~al.(1997)Costanza, d'Arge, de~Groot, Farber, Grasso,
  Hannon, Limburg, Naeem, ONeill, Paruelo, Raskin, Sutton, and van~den
  Belt}]{Costanza1997}
Costanza, R., d'Arge, R., de~Groot, R., Farber, S., Grasso, M., Hannon, B.,
  Limburg, K., Naeem, S., ONeill, R.~V., Paruelo, J., Raskin, R.~G., Sutton,
  P., van~den Belt, M., 1997. The value of the world's ecosystem services and
  natural capital. Nature 387~(6630), 253--260.

\bibitem[{de~Langre(2008)}]{deLangre2008}
de~Langre, E., 2008. Effects of wind on plants 40, 141--168.

\bibitem[{Denny et~al.(1998)Denny, Gaylord, Helmuth, and Daniel}]{Denny1998}
Denny, M., Gaylord, B., Helmuth, B., Daniel, T., 1998. The menace of momentum:
  Dynamic forces on flexible organisms. Limnology and Oceanography 43~(5),
  955--968.

\bibitem[{Ellington(1991)}]{Ellington1991}
Ellington, C.~P., 1991. Aerodynamics and the origin of insect flight. Adv.
  Insect Physiol 23, 171--210.

\bibitem[{Fonseca and Cahalan(1992)}]{Fonseca1992}
Fonseca, M.~S., Cahalan, J.~A., 1992. A preliminary evaluation of wave
  attenuation by four species of seagrass. Estuarine, Coastal and Shelf Science
  35~(6), 565--576.

\bibitem[{Fonseca et~al.(2007)Fonseca, Koehl, and Kopp}]{Fonseca2007}
Fonseca, M.~S., Koehl, M. A.~R., Kopp, B.~S., 2007. Biomechanical factors
  contributing to self-organization in seagrass landscapes. Journal of
  experimental marine biology and ecology 340~(2), 227--246.

\bibitem[{Ghisalberti and Nepf(2002)}]{Ghisalberti2002}
Ghisalberti, M., Nepf, H.~M., 2002. Mixing layers and coherent structures in
  vegetated aquatic flows. Journal of Geophysical Research-Oceans 107~(C2),
  3011.

\bibitem[{Gosselin et~al.(2010)Gosselin, de~Langre, and
  Machado-Almeida}]{Gosselin2010}
Gosselin, F., de~Langre, E., Machado-Almeida, B.~A., 2010. Drag reduction of
  flexible plates by reconfiguration. Journal of Fluid Mechanics 650, 319--341.

\bibitem[{Graham(1980)}]{Graham1980}
Graham, J. M.~R., 1980. The forces on sharp-edged cylinders in oscillatory flow
  at low keulegan-carpenter numbers. Journal of Fluid mechanics 97~(3),
  331--346.

\bibitem[{Huang et~al.(2011)Huang, Rominger, and Nepf}]{Huang2011}
Huang, I., Rominger, J., Nepf, H., 2011. The motion of kelp blades and the
  surface renewal model. Limnology and Oceanography 56~(4), 1453--1462.

\bibitem[{Hurd(2000)}]{Hurd2000}
Hurd, C.~L., 2000. Water motion, marine macroalgal physiology, and production.
  Journal of Phycology 36~(3), 453--472.

\bibitem[{Infantes et~al.(2012)Infantes, Orfila, Simarro, Terrados, Luhar, and
  Nepf}]{Infantes2012}
Infantes, E., Orfila, A., Simarro, G., Terrados, J., Luhar, M., Nepf, H., 2012.
  Effect of a seagrass (posidonia oceanica) meadow on wave propagation. Marine
  Ecology - Progress Series 456, 63--72.

\bibitem[{Keulegan and Carpenter(1958)}]{Keulegan1958}
Keulegan, G.~H., Carpenter, L.~H., 1958. Forces on cylinders and plates in an
  oscillating fluid. Journal of Research of the National Bureau of Standards
  60~(5), 423--440.

\bibitem[{Kobayashi et~al.(1993)Kobayashi, Raichle, and Asano}]{Kobayashi1993}
Kobayashi, N., Raichle, A.~W., Asano, T., 1993. Wave attenuation by vegetation.
  Journal of Waterway, Port, Coastal, and Ocean Engineering - ASCE 119~(1),
  30--48.

\bibitem[{Kundu and Cohen(2004)}]{Kundu2004}
Kundu, P.~K., Cohen, I.~M., 2004. Fluid Mechanics.

\bibitem[{Lowe et~al.(2005)Lowe, Koseff, and Monismith}]{Lowe2005}
Lowe, R.~J., Koseff, J.~R., Monismith, S.~G., 2005. Oscillatory flow through
  submerged canopies: 1. velocity structure. Journal of Geophysical Research C:
  Oceans 110~(10), 1--17.

\bibitem[{Luhar(2012)}]{LuharPhD}
Luhar, M., 2012. Analytical and experimental studies of plant-flow interaction
  at multiple scales. Ph.D. thesis, Massachusetts Institute of Technology.

\bibitem[{Luhar et~al.(2010)Luhar, Coutu, Infantes, Fox, and Nepf}]{Luhar2010}
Luhar, M., Coutu, S., Infantes, E., Fox, S., Nepf, H., 2010. Wave-induced
  velocities inside a model seagrass bed. Journal of Geophysical Research C:
  Oceans 115~(12).

\bibitem[{Luhar et~al.(2013)Luhar, Infantes, Orfila, Terrados, and
  Nepf}]{Luhar2013}
Luhar, M., Infantes, E., Orfila, A., Terrados, J., Nepf, H., 2013. Field
  observations of wave-induced streaming through a submerged seagrass
  (posidonia oceanica) meadow. Journal of Geophysical Research C: Oceans
  118~(4), 1955--1968.

\bibitem[{Luhar and Nepf(2011)}]{Luhar2011}
Luhar, M., Nepf, H.~M., 2011. Flow-induced reconfiguration of buoyant and
  flexible aquatic vegetation. Limnology and Oceanography 56~(6), 2003--2017.

\bibitem[{Madsen(1971)}]{Madsen1971}
Madsen, O.~S., 1971. On the generation of long waves. J.Geophys.Res. 76~(36),
  8672--8683.

\bibitem[{Manca et~al.(2012)Manca, Caceres, Alsina, Stratigaki, Townend, and
  Amos}]{Manca2012}
Manca, E., Caceres, I., Alsina, J., Stratigaki, V., Townend, I., Amos, C.,
  2012. Wave energy and wave-induced flow reduction by full-scale model
  posidonia oceanica seagrass. Continental Shelf Research 50-51, 100--116.

\bibitem[{Mass et~al.(2010)Mass, Genin, Shavit, Grinstein, and
  Tchernov}]{Mass2010}
Mass, T., Genin, A., Shavit, U., Grinstein, M., Tchernov, D., 2010. Flow
  enhances photosynthesis in marine benthic autotrophs by increasing the efflux
  of oxygen from the organism to the water. Proceedings of the National Academy
  of Sciences of the United States of America 107~(6), 2527--2531.

\bibitem[{Mei et~al.(2005)Mei, Stiassnie, and Yue}]{Mei2005}
Mei, C.~C., Stiassnie, M., Yue, D. K.~P., 2005. Theory and Applications of
  Ocean Surface Waves.

\bibitem[{Mendez and Losada(2004)}]{Mendez2004}
Mendez, F.~J., Losada, I.~J., 2004. An empirical model to estimate the
  propagation of random breaking and nonbreaking waves over vegetation fields.
  Coastal Engineering 51~(2), 103--118.

\bibitem[{Mendez et~al.(1999)Mendez, Losada, and Losada}]{Mendez1999}
Mendez, F.~J., Losada, I.~J., Losada, M.~A., 1999. Hydrodynamics induced by
  wind waves in a vegetation field. Journal of Geophysical Research C: Oceans
  104~(C8), 18383--18396.

\bibitem[{Mullarney and Henderson(2010)}]{Mullarney2010}
Mullarney, J.~C., Henderson, S.~M., 2010. Wave-forced motion of submerged
  single-stem vegetation. Journal of Geophysical Research C: Oceans
  115~(C12061).

\bibitem[{Sarpkaya and O'Keefe(1996)}]{Sarpkaya1996}
Sarpkaya, T., O'Keefe, J.~L., 1996. Oscillating flow about two- and
  three-dimensional bilge keels. Journal of Offshore Mechanics and Arctic
  Engineering 118, 1--6.

\bibitem[{Temmerman et~al.(2013)Temmerman, Meire, Bouma, Herman, Ysebaer, and
  Vriend}]{Temmerman2013}
Temmerman, S., Meire, P., Bouma, T.~J., Herman, P. M.~J., Ysebaer, T., Vriend,
  H. J.~D., 2013. Ecosystem-based coastal defence in the face of global change.
  Nature 504, 79--83.

\bibitem[{Vogel(1994)}]{Vogel1994}
Vogel, S., 1994. Life in moving fluids, 2nd Edition. Princeton Univ. Press,
  Princeton, NJ.

\bibitem[{Zeller et~al.(2014)Zeller, Weitzman, Abbett, Zarama, Fringer, and
  Koseff}]{Zeller2014}
Zeller, R.~B., Weitzman, J.~S., Abbett, M.~E., Zarama, F.~J., Fringer, O.~B.,
  Koseff, J.~R., 2014. Improved parameterization of seagrass blade dynamics and
  wave attenuation based on numerical and laboratory experiments. Limnology and
  Oceanography 59~(1), 251--266.

\bibitem[{Zimmerman(2003)}]{Zimmerman2003}
Zimmerman, R.~C., 2003. A biooptical model of irradiance distribution and
  photosynthesis in seagrass canopies. Limnology and Oceanography 48~(1),
  568--585.

\end{thebibliography}
\end{document}